\documentclass[prl,aps,twocolumn,showpacs,superscriptaddress,longbibliography]{revtex4-1}
\usepackage{cleveref}
\usepackage{graphicx}% Include figure files
\usepackage{dcolumn}% Align table columns on decimal point
\usepackage{bm}% bold math
\usepackage{amssymb,amsmath}
\usepackage{sidecap}
\usepackage{wrapfig}
\usepackage{natbib}
\usepackage{dsfont}
 \usepackage{makecell} % to break lines inside a table cellx

\usepackage{graphicx}
\usepackage{leftidx}
\usepackage{color}

\usepackage{bbold} %This package allows to write the identity symbol

\usepackage{wasysym} % In order to introduce polygon symbols in the text
\usepackage{appendix}

\usepackage{stackrel}

\usepackage{soul,xcolor}
\setstcolor{red}

 \newcommand{\rmi}{{\rm{i}}}
\newcommand{\be}{\begin{equation}}
\newcommand{\ee}{\end{equation}}
\newcommand{\ba}{\begin{align}}
\newcommand{\ea}{\end{align}}
\newcommand{\sysb}{\left\{\begin{array}}
\newcommand{\syse}{\end{array}\right.}
\newcommand{\baa}{\begin{array}}
\newcommand{\eaa}{\end{array}}
\newcommand{\bs}{\begin{split}}
\newcommand{\es}{\end{split}}

\newcommand{\matb}{\left(\begin{array}}
\newcommand{\mate}{\end{array}\right)}

\newcommand{\mal}{\mathcal}
\newcommand{\rmd}{{\rm{d}}}

\newcommand{\rme}[1]{{\rm{e}}^{#1}}
\newcommand{\mand}{\quad\text{ and }\quad}

\newcommand{\wt}{\widetilde}

\newcommand{\trace}[1]{{\rm tr}\left\{ #1 \right\}}

\newcommand{\ha}{\frac{1}{2}}

\newcommand{\lt}{\left(}
\newcommand{\rt}{\right)}
\newcommand{\lqq}{\left[}
\newcommand{\rqq}{\right]}
\newcommand{\lan}{\left\langle}
\newcommand{\ran}{\right\rangle}
\newcommand{\abs}[1]{\left| #1 \right|}

\newcommand{\av}[1]{\lan #1 \ran}

\newcommand{\set}[1]{\left\{  #1  \right\}}

\newcommand{\R}{\mathbb{R}}

\newcommand{\Z}{\mathbb{Z}}

\newcommand{\ket}[1]{\left| #1 \ran}
\newcommand{\bra}[1]{\lan #1 \right|}

\newcommand{\proj}[1]{\ket{#1} \bra{#1}}
\newcommand{\comm}[2]{\left[ #1, #2 \right]}

\newcommand{\cosa}[1]{\cos \left(  #1 \right)}
\newcommand{\sina}[1]{\sin \left(  #1 \right)}

\newcommand{\cosha}[1]{\cosh \left(  #1 \right)}
\newcommand{\sinha}[1]{\sinh \left(  #1 \right)}

\newcommand{\prodl}[2]{\prod\limits_{#1}^{#2}}

\newcommand{\change}[1]{\textcolor{black}{#1}}
\newcommand{\changer}[1]{\textcolor{black}{#1}}

\usepackage{amsthm}

 \newcommand{\up}{\uparrow}
 
 \newcommand{\down}{\downarrow}

  \newcommand{\op}[1]{\mathrm{\hat{#1}}}

\begin{document}

\title{Synthetic lattices, flat bands and localization in Rydberg quantum simulators}

\author{Maike Ostmann}
\author{Matteo Marcuzzi}
\author{Ji\v{r}\'{i} Min\'{a}\v{r}}
\author{Igor Lesanovsky}
\affiliation{School of Physics and Astronomy, University of Nottingham, Nottingham, NG7 2RD, UK}
\affiliation{Centre for the Mathematics and Theoretical Physics of Quantum Non-equilibrium Systems,
University of Nottingham, Nottingham NG7 2RD, UK}

\begin{abstract}
The most recent manifestation of cold Rydberg atom quantum simulators that employs tailored optical tweezer arrays enables the study of many-body dynamics under so-called facilitation conditions. We show how the facilitation mechanism yields a Hilbert space structure in which the many-body states organize into synthetic lattices, which feature in general one or several flat bands and may support immobile localized states. We focus our discussion on the case of a ladder lattice geometry for which we analyze in particular the influence of disorder generated by the uncertainty of the atomic positions. The localization properties of this system are characterized through two localization lengths which are found to display anomalous scaling behavior at certain energies. Moreover, we discuss the experimental preparation of an immobile localized state, and analyze disorder-induced propagation effects.
\end{abstract}
\maketitle

Over the past few decades, advances in the manipulation of cold and ultra-cold atomic gases rendered them into a versatile quantum simulation platform \cite{Bloch_2008,Bloch_2012}. Indeed, several paradigmatic many-body models have been studied experimentally, including the Luttinger liquid \cite{hofferberth2007}, the Tonks-Girardeau gas \cite{kinoshita2004} as well as Bose-Hubbard \cite{greiner2002, greiner2003} and Fermi-Hubbard Hamiltonians \cite{Kohl2005}, permitting to directly observe several predicted phenomena, such as quantum revivals \cite{greiner2002_revival}, Lieb-Robinson bounds \cite{cheneau2012}, and topological phase transitions \cite{hadzibabic2006}.

Among many different physical systems apt to act as quantum simulators, ensembles of Rydberg atoms \cite{a_Saffman_RMP_10, Low_2012, Gallagher_1994} stand out for their strong interactions, which are now known to give rise to an intricate phenomenology, including devil's staircases \cite{Weimer2010, Levi2016, Lan2015}, aggregate formation and melting \cite{Schempp2014, Lan2016}, Rydberg crystals \cite{Schauss_2015}, optical bistability \cite{Carr2013, Sibalic2016} and phase transitions or universal scaling \cite{Low2009, Marcuzzi2014, Gutierrez2015}. These systems are currently employed for a variety of tasks, such as quantum information processing \cite{Jaksch2000,Weimer_2010,Saffman_2016} and the simulation of quantum spin systems \cite{Labuhn_2015, Schauss_2015}. Several among these instances employ the so-called \emph{facilitation} (or \emph{anti-blockade}) mechanism (see e.g., Refs.~\cite{Ates_2007,Amthor_2010,Garttner_2013,schonleber2014,Lesanovsky_2014,Urvoy_2015,Valado_2016}) to actuate a form of quantum transport.

In quantum systems, it is well-established that transport can be heavily affected by the presence of quenched disorder, a phenomenon known as Anderson localization \cite{Anderson1958}. In the presence of randomly-distributed impurities in a metal, for example, different paths taken by an electron can interfere destructively, leading to localization. In one and two dimensions, this effect is so relevant that for arbitrarily small disorder all wavefunctions are localized and transport is effectively impossible \cite{Mott1961,Ishii1973}. Since their first prediction, these effects have been experimentally observed in a range of systems, spanning electron gases \cite{Cutler:1969}, cold atoms \cite{Billy:2008,Roati:2008,Semeghini:2015}, thin films \cite{Liao:2015} or periodically-driven nitrogen molecules \cite{Bitter:2016}.

Apart from the case of quenched disorder, localized states can also arise in tight-binding models from particular lattice geometries. In these cases, destructive interference leads to the emergence of flat bands. Models with flat bands typically allow the construction of localized eigenstates, and have been experimentally realized with cold atoms \cite{Shen2010}, photonic lattices \cite{Mukherjee2015}, and synthetic solid-state structures \cite{slot2017, drost2017}. When disorder is introduced in such systems, these pre-existing localized states couple to the dispersive, system-spanning ones and start acting like scatterers, inducing a richer phenomenology, such as localization enhancement \cite{Leykam2017}, Anderson transitions in lower-dimensional systems \cite{Bodyfelt2014}, and disorder-induced delocalization \cite{Goda2006}.

In this paper we demonstrate that Rydberg lattice quantum simulators \cite{Schauss_2015,Labuhn_2015,Bernien2017} permit the exploration of disorder phenomena in the presence of flat bands. We show that under facilitation conditions -- when the system parameters are set such that Rydberg states can only be excited next to an already existing excitation -- the Hilbert space acquires a regular lattice structure featuring flat bands. In this picture, the uncertainty of atomic positions translates into a disordered potential on the newly created synthetic lattice. Scenarios similar to these were previously theoretically analyzed in \cite{Leykam2017, Bodyfelt2014}. Here we show that they emerge naturally in Rydberg quantum simulators employing optical tweezer arrays \cite{Labuhn_2015,Jau2016,Bernien2017}. We illustrate our findings for the case of a so-called ``Lieb ladder'': we analyze the scaling of the localization length and discuss the spreading dynamics of a local flat-band eigenstate under the action of different disorder strengths.

\emph{Facilitation, Hilbert space structure and flat bands---} We start by considering a regular \cite{footnote1} lattice of $N$ optical tweezers, each loaded with a single Rydberg atom, and with nearest-neighbor distance $R_0$. A laser is shone with a frequency detuned by $\Delta$ with respect to an atomic transition between the electronic ground state $\ket{\down}$ and a Rydberg level $\ket{\up}$. We work here in natural units $\hbar = 1$. Atoms in the Rydberg state $\ket{\up}$ interact, at distance $d$, via an algebraically-decaying potential $V(d) = C_\alpha / d^\alpha$, with $\alpha = 3 (6)$ for dipole-dipole (van-der-Waals) interactions (without loss of generality we choose $C_\alpha > 0$). Within the rotating wave approximation the Hamiltonian of this system reads
\begin{align}
 \op{H} = \Omega \, \sum_{k=1}^N  \op{\sigma}_x^{(k)} \, + \, \Delta\, \sum_{k=1}^N\,\op{n}_k +\,  \,
\frac{1}{2} \sum_{\substack{k= 1\\ m \ne k}}^N \, V(d_{km}) \, \op{n}_m\, \op{n}_k,
 \label{Eq:Hamil_full}
\end{align}
where $\Omega$ is the laser Rabi frequency, $k$ and $m$ are lattice indices, $d_{km}$ denotes the distance between atoms in sites $k$ and $m$, $\op{\sigma}_x^{(k)} = \ket{\up_k} \bra{\down_k} + \ket{\down_k} \bra{\up_k}$ and $\op{n}_k = \proj{\up_k}$. The facilitation condition is obtained by setting $\Delta = -V(R_0)$, so that an isolated excited atom makes the transitions of its neighbors resonant with the laser. In the following, we consider $\abs{\Delta} \gg \Omega$, so that non-facilitated atoms are sufficiently off-resonant to neglect their excitation. Furthermore, we require $V(2R_0) \gg \Omega$ which still ensures that an isolated excitation can facilitate the production of another on a neighboring site, but suppresses the creation of additional excitations in the neighborhood. For example, in one dimension $\abs{\bra{\up \up \up} \rme{-iHt} \ket{\up \up \down}}^2 \sim O((\Omega / V(R_1))^2)$. In the following, we neglect these strongly suppressed transitions, effectively splitting the Hilbert space into subspaces separated by energy scales $\gg \Omega$. Each subspace comprises a set of quasi-resonant states separated by scales $\sim O(\Omega)$ (see Ref.~\cite{a_Marcuzzi_PRL_17} for more details on this structure). Intuitively, this means that a single excitation can at most produce one more in the neighborhood, after which either the former facilitates the de-excitation of the latter, or vice versa.

Amidst all various subspaces, the simplest non-trivial choice corresponds to the one consisting of all states with either a single excitation or a single pair of excitations on neighboring sites \cite{Mattioli2015, a_Marcuzzi_PRL_17}. Hence, as sketched in Fig.~\ref{Fig:flat_band_lattices} for a few planar examples, a lattice structure emerges in the Hilbert space which closely resembles the real-space geometry of the tweezer arrays. These synthetic lattices are constructed via the following rules:
\begin{figure}
\includegraphics[width=\columnwidth]{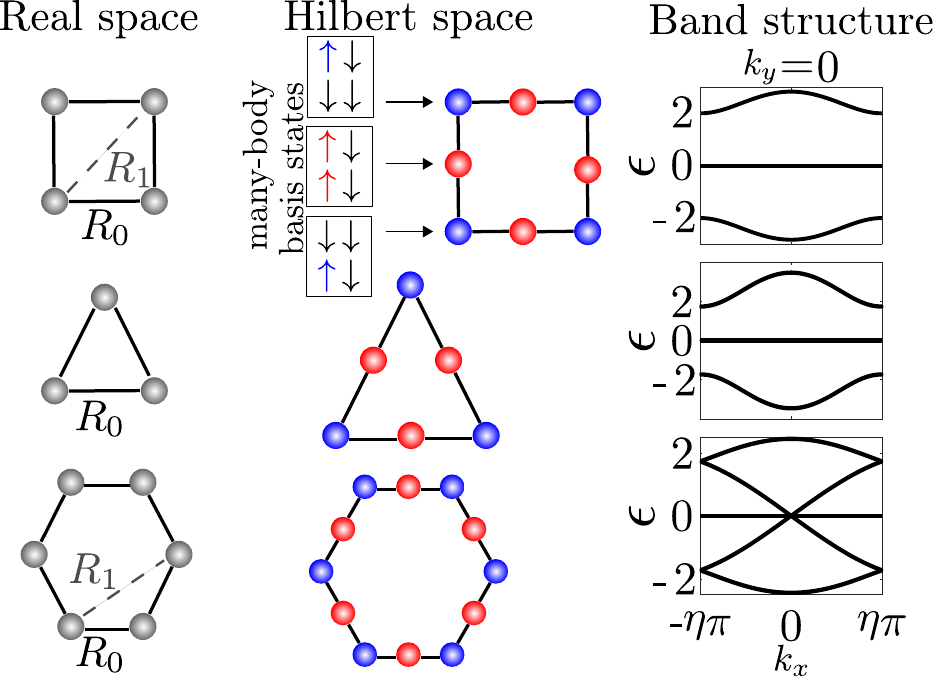}
\caption{\textbf{Left column:} geometry of a square, a triangular, and a honeycomb lattice in real space. The gray dots depict the position of the Rydberg atoms and the lines the interaction between neighboring atoms. $R_0$ and $R_1$ represent nearest and next-nearest neighbor distances, respectively. \textbf{Middle column:} respective "synthetic lattices" in the Hilbert space under facilitation conditions. The blue dots represent one-excitation states while the red ones are pair states. \textbf{Right column:} Cut through the Brillouin zone for each lattice geometries at $k_y= 0$. Each lattice features (at least) a flat band. The momentum scales for the three lattices (from top to bottom) are $\eta=(1,\tfrac{4}{3},\tfrac{4}{3}).$}
\label{Fig:flat_band_lattices}
\end{figure}
(i) in the original lattice structure, draw the links joining nearest neighbors; (ii) identify each site with the state having a single excitation on that site. This exhausts all ``one-excitation'' states in the subspace; (iii) each ``pair'' state can be straightforwardly associated to the link joining the two excited atoms; hence, place an additional site in the midpoint of each link and associate it with the corresponding ``pair'' state. The links in this new-found structure now effectively represent a pair of states connected by the Hamiltonian, which can be therefore seen as a tight-binding model on a generalized synthetic lattice. In the case of a square lattice, the new structure (see Fig.~\ref{Fig:flat_band_lattices}) is the \emph{Lieb lattice} and is known to feature one flat and two dispersive bands which meet with a linear dispersion at the edges of the first Brillouin zone. However, this construction is general and can be extended to any kind of regular \cite{footnote1} lattice. Most of these structures will support flat bands as well: It can be shown \cite{SM} that, calling $n_1$ ($n_2$) the number of one-excitation (pair) states in a unit cell, the number of flat bands $n_{\rm flat}$ must be $\geq \abs{n_1 - n_2}$. For the examples of Fig.~\ref{Fig:flat_band_lattices}, the square, triangular and honeycomb lattices have $(n_1,n_2,n_{\rm flat}) = (1,2,1)$, $(1,3,2)$ and $(2,3,1)$ respectively. These flat bands constitute extensively-degenerate eigenspaces of the Hamiltonian; as such, it is often possible to recombine the usual (plane-wave-like) Bloch solutions to form a set of localized (or immobile) eigenstates.

\emph{Disorder---} Disorder enters the picture through the uncertainty in the atomic positions. Even small displacements from the center of the traps can significantly shift the atomic transitions off resonance from the laser frequency, thereby hindering the facilitation mechanism \cite{a_Marcuzzi_PRL_17}. In fact, the interaction potential seen by an atom at a distance $R = R_0 + \delta R$ from an excitation will be $V(R) = V(R_0 + \delta R) \equiv V(R_0) + \delta V$. At small disorder ($\delta R \ll R_0$ and $\delta V \ll V(2R_0)$) the energy shifts can be approximated by $\delta V \approx -\alpha C_\alpha / R_0^{\alpha + 1} \delta R$ \cite{SM}. These random variables only affect pair states, creating a disordered potential landscape over the pair (red) sites in Fig.~\ref{Fig:flat_band_lattices}.

In order to characterize the disorder, we denote by $\omega$ the optical tweezer trapping frequency (assumed hereafter to be isotropic in space), by $m$ the atomic mass and by $T$ the temperature. The probability distribution of a trapped atom can then be approximately described as a Gaussian of width $\sigma$ around the trap center. We require now that (I) $k_B T \gg \hbar \omega$: this implies that one can use the semiclassical estimate $\sigma \approx \sqrt{k_B T / m\omega^2}$ and moreover that the thermal de Broglie wavelength of the atom is much smaller than the distribution width. In other words, the atom can be approximately considered localized somewhere within the trap according to a classical probability distribution. (II) $\omega \Delta t \ll 1$, with $\Delta t$ the duration of an experiment: this ensures that the atoms will not appreciably move from their positions in this time frame and thus the disorder is quenched. (III) $\Omega \gg \omega$, or in other words the dynamics of the internal degrees of freedom is much faster than the one of the kinetic ones, so that within an experiment one can probe the action of the disordered Hamiltonian on the system while keeping the specific realization of the disorder fixed. The properties of the probability distribution of energy shifts are discussed in \cite{SM}; here we just mention that amplitudes of shifts over different pair sites are not independent, but correlated.

\begin{figure}
\includegraphics[width=\columnwidth]{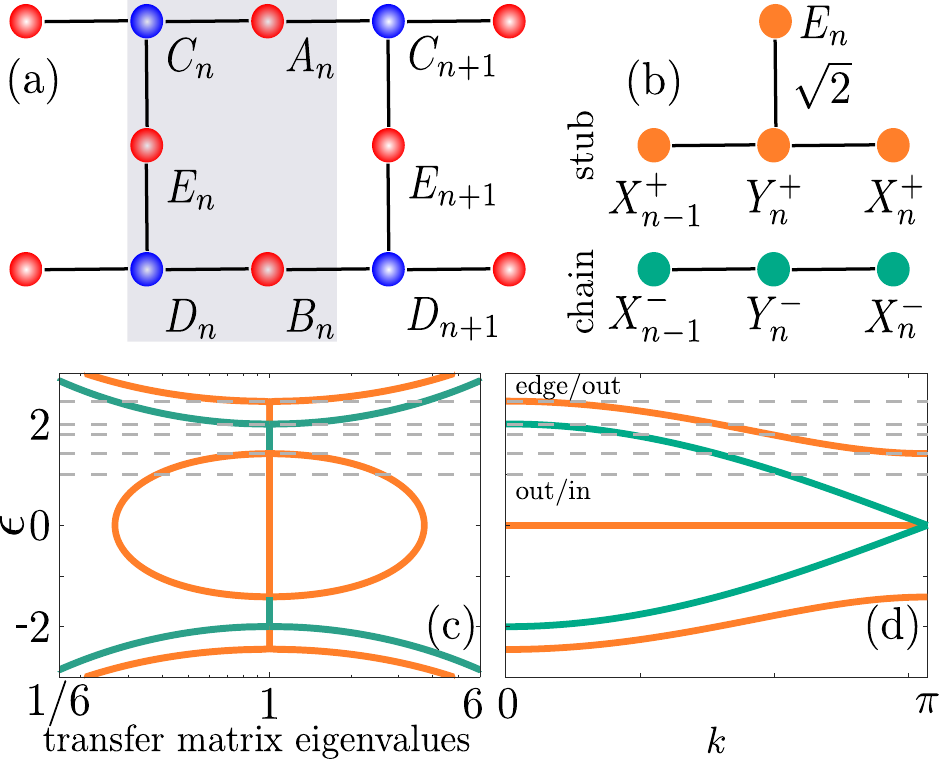}
\caption{Hilbert space structure and spectrum in the absence of disorder. \textbf{(a)} Lieb ladder; blue (red) dots correspond to one-excitation (pair) states. We introduce a convenient notation for the five sites $A_n$, $B_n$, $C_n$, $D_n$, $E_n$ in the $n$-th unit cell (shaded gray). \textbf{(b)} A change of basis -- the so-called ``detangling'', introducing the new linear combinations $X_n^\pm = (A_n \pm B_n)/\sqrt{2}$ and $Y_n^\pm = (C_n \pm D_n)/\sqrt{2}$ \cite{a_Flach_EPL_14,Leykam2017} -- maps the Lieb ladder onto two decoupled chains. The $\sqrt{2}$ factor denotes that the hopping amplitude on the vertical link of each unit cell is amplified by that same amount. \textbf{(c)} Eigenvalues of the transfer matrix in log-linear scale. The dotted lines corresponds to the energies $\epsilon = \{1, \sqrt 2, 1.8, 2, \sqrt 6\}$ at which the scaling of the localization lengths is investigated in Fig.~\ref{Fig:2D_loc_length}. \textbf{(d)} Band structure of the Lieb ladder. The bands corresponding to the stub lattice are given in orange and bands of the ordinary 1D chain are shown in green.}
\label{Fig:decoupling}
\end{figure}

\emph{Disordered Lieb ladder---} In the remainder of our discussion, we shall focus on a ladder configuration, i.e.~a quasi-one-dimensional lattice formed by placing two linear chains parallel to each other at a lattice spacing $R_0$. For this example, the synthetic lattice (a ``1D Lieb lattice'') in the Hilbert space is sketched in Fig.~\ref{Fig:decoupling}(a). The unit cell consists of five sites with $n_1 = 2$ and $n_2 = 3$ and the band structure features one zero-energy flat and four dispersive bands [Fig.~\ref{Fig:decoupling}(d)].

This Lieb ladder constitutes one of the simplest examples where flat bands produce a non-trivial interplay with the on-site disorder \cite{Leykam2017}. In a Rydberg quantum simulator, however, the disorder only appears on pair states, i.e. all the blue sites of the synthetic lattice [Fig.~\ref{Fig:flat_band_lattices}(a)] are unaffected by it. To investigate the effect of this unusual disorder scenario we study in the following the scaling behavior of the \emph{localization length} $\xi$ for small disorder strengths. This quantity encodes the localization properties of the energy eigenstates, whose amplitude is typically peaked in a specific area of the lattice and decays exponentially as $\rme{-r/\xi}$ at large distances $r$.

For a ladder like the one under study, two different values of $\xi$ can be extracted at any given energy, which we denote by $\xi_{1/2}$ and order according to $\xi_1 < \xi_2$. To elucidate the reason, one can perform an appropriate change of basis (``detangling transformation'' \cite{a_Flach_EPL_14,Leykam2017}) through which the Lieb ladder is mapped onto two uncoupled one-dimensional lattices [see Fig.~\ref{Fig:decoupling}(b)], a chain (in green, supporting the two innermost dispersive bands) and a stub lattice (in orange, supporting the flat and two outermost dispersive bands) \cite{SM}. At small disorder, one can thus associate either localization length to one of the two detangled chains.

\begin{figure}
\includegraphics[width=\columnwidth]{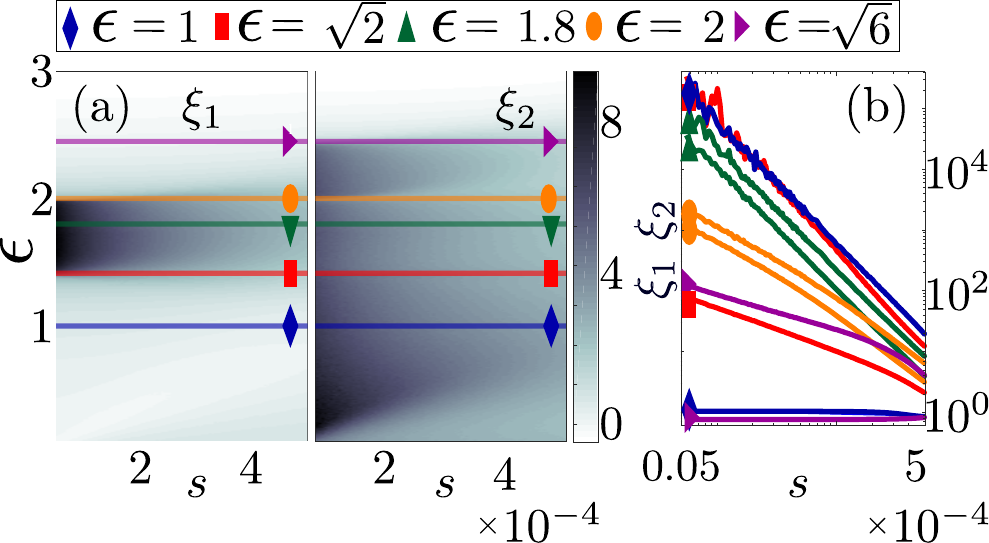}
\caption{\textbf{(a)} Localization lengths $\xi_1,\,\xi_2$ as a function of the energy $\epsilon$ and the disorder strength $s = \sigma / R_0$.
\textbf{(b)} Localization lengths along each of the solid lines displayed in panel (a) in log-log scale. For small disorder all lines are approximately linear which allows to assign approximate power law exponents $\nu$ characterizing the small-disorder behavior $\xi_i \sim s^\nu$: grouping them by energy $\epsilon$, they read $\nu\left(\epsilon = 1\right) \approx$ \{0, 2.2\}, $\nu\left(\epsilon = \sqrt{2}\right)\approx$ \{0.7, 2.2\}, $\nu\left(\epsilon = 1.8\right)\approx$ \{2.0, 1.9\}, $\nu\left(\epsilon = 2\right)\approx$ \{1.1, 1.1\}, $\nu\left(\epsilon = \sqrt{6}\right)\approx$ \{0, 0.6\}. For these computations we chose a dipole-dipole interaction ($\alpha = 3$) with an interaction strength of $V(R_0) = 300\Omega$. \changer{It is apparent that the lowermost curves bend down in the rightmost part of panel (b). Data in this region have been discarded to extract the slope.}}
 \label{Fig:2D_loc_length}
\end{figure}

The localization length $\xi_{1/2}$ are found numerically via a transfer matrix formalism and are displayed in Fig.~\ref{Fig:2D_loc_length}(a) as a function of the disorder strength $s \equiv \sigma / R_0$ and the energy $\epsilon$.
In Fig.~\ref{Fig:2D_loc_length}(b) we display log-log plots of the correlation lengths at selected energies as functions of $s$, which illustrate algebraic scaling $\xi_i \sim s^\nu$, for sufficiently small $s$. Where possible, we connect our findings to those presented in Ref.~\cite{Leykam2017}, where the same geometry is studied with independent disorder on all sites. The usual scaling for Anderson localization corresponds to $\nu = 0$ at energies outside a band (``out''), $\nu = 2/3$ at a band edge (``edge'') and $\nu = 2$ inside a band (``in''). The energies selected in Fig.~\ref{Fig:decoupling} correspond to $\epsilon =1$ (out/in), $\sqrt{2}$ (edge/in), $1.8$ (in/in), $2$ (in/edge) and $\sqrt{6}$ (edge/out). Here the entries in the brackets refer to the two band structures depicted in Fig.~\ref{Fig:decoupling}(c,d): (orange/green).
6
In Ref.~\cite{Leykam2017} an ``anomalous'' scaling $\nu = 4/3$ was found at $\epsilon = \sqrt{2}$ and $2$. This was attributed to the fact that disorder, in the detangled picture, is not merely on-site but couples the two chains. This in turn may produce resonances between states in the middle of a band and states at the edge of the other when the latter displays vanishing group velocity. \changer{Comparing these values with the ones obtained for our situation, we observe reasonable agreement at $\epsilon = 1$, $\epsilon = 1.8$ and $\epsilon = \sqrt{6}$, plus for the ``edge'' scaling at $\epsilon = \sqrt{2}$. The anomalous ``in'' scaling at $\epsilon = \sqrt{2}$ seems instead to be ``cured'' as we retrieve a result compatible with the usual Anderson one ($\nu  \approx 2$). As we show in \cite{SM}, this is likely to be due to the alternating structure of the disorder in the synthetic lattice, which in the detangled picture results in the absence of random couplings between $Y_n^{\pm}$ sites [see Fig. \ref{Fig:decoupling}(b)], present instead in Ref. \cite{Leykam2017}.}

\changer{We find however discrepancies at $\epsilon = 2$, where both localization lengths are close to $1.1$ and do not seem to match with either of the expected values $2/3$ (edge) or $4/3$ (in, anomalous). An explanation for this behavior, which does not seem to be related simply to the alternating structure of the disorder \cite{SM}, is currently lacking and requires further investigations.}

\emph{Localized flat band state dynamics---}
\begin{figure}
\includegraphics[width=\columnwidth]{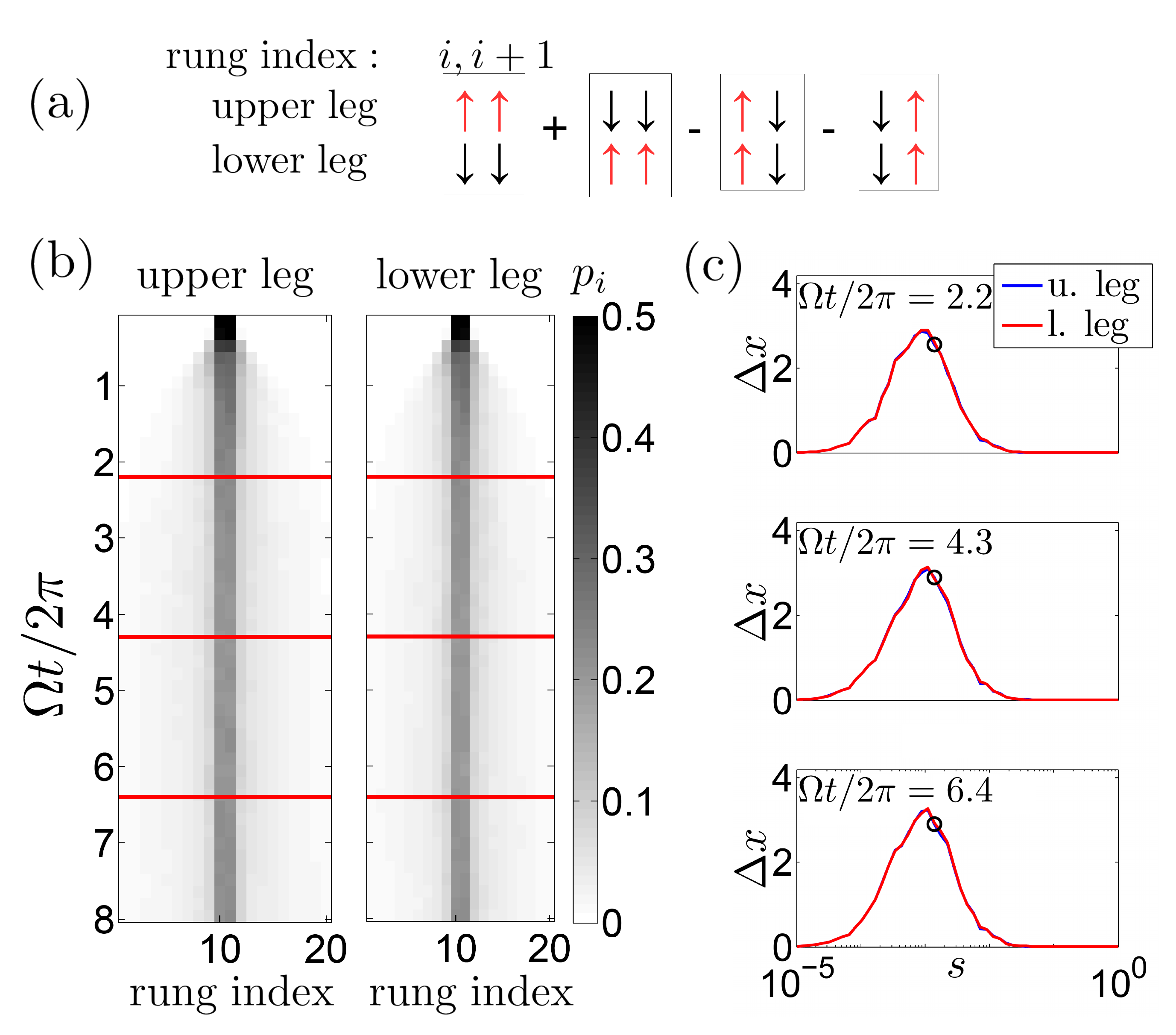}
\caption{\textbf{(a)} Schematic representation of the spin configuration corresponding to the initial state $\ket{\psi_{\rm loc}}$ localized at rungs $i,i+1$ of the ladder. \textbf{(b)} The probability of excitations $p_i$ given by the time evolution of the localized state with an initial support in the middle (rungs $10$ and $11$) of the ladder of length $20$ for $s=0.0014$. The left (right) pane shows the time evolution in the upper (lower) leg of the ladder. The horizontal red lines denote three different times for which the respective value of $\Delta x$ is shown as a black circle in (c). \textbf{(c)} Standard deviation of the excitation positions $\Delta x$ vs. the disorder strength $s$ for three different times. Blue (red) solid lines, which are virtually indistinguishable correspond to upper (lower) leg of the ladder respectively. Results obtained for $100$ disorder realizations and $V(R_0)=200\Omega$.}
\label{Fig:time evolution}
\end{figure}
Experimentally measuring the localization lengths studied above is challenging due to the required large systems size and small disorder amplitudes. However, one can probe the influence of disorder by initializing the system in a specific state and tracking the subsequent dynamics by measuring the on-site excitation probabilities \cite{Schauss_2015,Labuhn_2015,Bernien2017}. A particularly interesting choice for an initial state is localized and an eigenstate of the flat band. Such state is not propagating in the absence of disorder. We show in \cite{SM} that it takes the form $\ket{\psi_{\rm loc}} = 1/\sqrt{4} \left( \ket{A_i} + \ket{B_i} - \ket{E_i} - \ket{E_{i+1}} \right)$, being entirely localized at rungs $i,i+1$ of the ladder [see Fig. \ref{Fig:time evolution}(a)]. States of this form can be prepared experimentally via single site addressing \cite{SM}.

The time evolution of the excitation density is shown in Fig. \ref{Fig:time evolution}(b). The effect of the disorder becomes apparent in the width $\Delta x$ \cite{SM} of the density packet which quickly reaches a stationary state. It is interesting to observe that, as shown in Fig.~\ref{Fig:time evolution}(c), the stationary value of $\Delta x$ shows a non-monotonic behavior as a function of $s$. This can be understood as follows: at very small (but finite) disorder strength $s$ the initial state (energy $\epsilon \approx 0$) is almost a flat band eigenstate and it therefore only minimally spreads (see e.g. Refs. \cite{PhysRevLett.114.245503,Goda2006}). As $s$ is increased, this picture breaks down and the state more and more strongly hybridizes with other states, allowing transport over larger distances. At the same time, however, the localization lengths at $\epsilon = 0$ decrease. Hence, an interplay ensues: the spreading $\Delta x$ of the state increases with $s$ as long as the localization length remains larger ($\Delta x \ll \xi_i$). Once the decrease in the localization scale catches up with the increase of $\Delta x$, the behavior is dominated by localization and, as expected, decreases with increasing disorder strength.

\emph{Conclusions and Outlook---} We have shown that Rydberg quantum simulators allow to explore localization phenomena in synthetic lattices with flat bands and unconventional types of disorder (correlated, alternating). The current study focuses on the Lieb ladder and on a particular excitation sector. Higher-dimensional lattices hosting more excitations are straight-forwardly realizable in experiment. It is thus a future theoretical challenge to shed light on these intricate and unexplored scenarios.

\emph{Acknowledgments---} The research  leading  to  these  results  has  received  funding  from the European Research Council under the European Union’s Seventh Framework Programme (FP/2007-2013)/ERC Grant Agreement No.
335266 (ESCQUMA), the EPSRC Grant No. EP/M014266/1, and the H2020-FETPROACT-2014 Grant No. 640378 (RYSQ). I.L. gratefully acknowledges funding through the Royal Society Wolfson Research Merit Award.

% \bibliographystyle{apsrev4-1}
% 
% \bibliography{bibliography}


\begin{appendix}
 
\onecolumngrid
 \begin{center}
{\Large{Supplemental Material: Synthetic lattices, flat bands and localization in Rydberg quantum simulator}s}
\end{center}
\twocolumngrid

\section{Approximate Gaussian distribution of the atoms} 
In order to show how the Gaussian distribution of the atomic positions arises, we consider here an atom of mass $m$ sitting in a one-dimensional optical trap of frequency $\omega$. The results will be straightforwardly generalizable to the three-dimensional case as the three Cartesian coordinates decouple and can be treated independently.
We work in a regime of temperatures $T$ much lower compared to the trap depth, but larger than the trap frequency, i.e., $k_B T \gg \hbar \omega$. The first assumption allows us to treat the trap as an harmonic potential, yielding a Hamiltonian
\be
	\op H_{\rm trap} \approx \frac{\hat{p}^2}{2m} + \frac{m}{2} \omega^2 \hat{x}^2,
\ee
where $\hat{x}$ and $\hat{p}$ are the quantum position and momentum \change{operators}, respectively. The thermal state of the system is described by the Gibbs form
\be
	\rho_{\rm th} = \frac{1}{Z} \rme{-\beta \op H_{\rm trap}},
\ee
where $\beta = 1/ (k_B T)$ and $Z$ is the partition function
\be
	Z = \trace{ \rme{-\beta \op H_{\rm trap}}}.
	\label{eq:part}
\ee
Employing the standard mapping
\be
	\hat{p} = \rmi \sqrt{\frac{\hbar m \omega}{2}} (\hat{b}^\dag - \hat{b}), \quad \hat{x} = \sqrt{\frac{\hbar }{2m \omega}} (\hat{b}^\dag + \hat{b})
\ee
in terms of bosonic creation ($\hat{b}^\dag $) and annihilation ($\hat{b} $) operators $\left(\comm{\hat{b}}{\hat{b}^\dag} = 1\right)$, one readily obtains
\begin{subequations}
\begin{align}
	&H = \hbar \omega \lt \hat{b}^\dag \hat{b} + \ha \rt  \\
	&\change{\mand}\notag\\
	&Z = \sum_n \rme{-\beta \hbar \omega (n + 1/2)} = \frac{1}{2 \sinha{\frac{\beta \hbar \omega}{2}}}\, .
\end{align}
\end{subequations}

Calling $\ket{x}$ the position eigenvector $\hat{x} \ket{x} = x \ket{x}$, the probability density functions of the atomic position is defined as
\be
	p_{\rm pos}(x) = \bra{x} \op \rho_{\rm th} \ket{x}.
\ee
Its analytical form can be extracted from the Feynman propagator for the harmonic oscillator $K(x,y,t) = \bra{x}  \rme{-\rmi t \op H / \hbar}   \ket{y}$, which, in the time interval $t \in (0, \pi/\omega)$, reads (see, e.g., \cite{kernel1, kernel2} for detailed derivations)
\be
\begin{split}
	K(x,y,t) = &\sqrt{\frac{m\omega}{2\pi \hbar \rmi \sina{\omega t}}}  \times \\
	\times & \exp \left\{  \rmi\frac{ m \omega}{2\hbar \sina{\omega t}} \lqq  (x^2 + y^2) \cosa{\omega t} - 2xy  \rqq  \right\}.
\end{split}
\ee 
Substituting $t \to -\rmi\beta \hbar$ and $y \to x$ one finds
\be
\begin{split}
	K(x,x,-\rmi\beta) & = \bra{x}  \rme{ - \beta H}   \ket{x} = \sqrt{\frac{m\omega}{2\pi \hbar \sinha{\omega \beta \hbar}}} \times \\
	\times & \exp \left\{  - \frac{ m \omega}{\hbar \sinha{\omega \beta \hbar}} \lt \cosha{\omega \beta \hbar} - 1  \rt x^2  \right\}.
\end{split}
\ee
Dividing by the partition function \eqref{eq:part} one finally finds the Gaussian
\be
\begin{split}
	p_{\rm pos}(x) & = \sqrt{\frac{m\omega ( \cosha{\omega \beta \hbar}-1) }{\pi \hbar \sinha{\omega \beta \hbar}}} \times \\
	\times & \exp \left\{  - \frac{ m \omega}{\hbar \sinha{\omega \beta \hbar}} \lt \cosha{\omega \beta \hbar} - 1  \rt x^2  \right\}.
	\label{eq:distr1d}
\end{split}
\ee
The variance $\sigma$ can be read off directly and amounts to
\be
	\sigma^2  = \frac{\hbar  \sinha{\omega \beta \hbar}}{ 2 m \omega (\cosha{\omega \beta \hbar}-1) }\,.
\ee
Since we assumed $\beta \ll \hbar \omega$, i.e., $\omega \beta \hbar \ll 1$, we can expand this expression to lowest order, which yields
\be
	\sigma^2 = \frac{1}{m \omega^2 \beta} = \frac{k_B T}{m \omega^2},
\ee
as reported in the main text. 

The distribution \eqref{eq:distr1d} is straightforwardly generalized to three-dimensions and traps centered along a \change{single linear} chain at positions $k\mathbf{R}_0 = (0,0,kR_0)$ with $k$ an integer:
\be
	p^{(k)}_{\rm pos}(\mathbf{r}) = \frac{1}{\lt 2\pi \rt^{3/2} \sigma_1 \sigma_2 \sigma_3} %\exp\left[-\sum_{i=1}^3\frac{(x_i - j r_i)^2}{2\sigma_i^2}. \right]
	\rme{-  \frac{r_1^2}{2\sigma_1^2} -\frac{r_2^2}{2\sigma_2^2} -\frac{(r_3 - (k-1) \cdot R_0)^2}{2\sigma_3^2} } .
	\label{eq:distr0}
\ee
For clarity, we remark here that the indices in the expression above distinguish between Cartesian components only, e.g, $r_1$ and $r_2$ are the components of the same atom along the $x$ and $y$ directions. In the following, when necessary the trap index will always appear before the component one, e.g., $r_{k,i}$ is the $i$-th component of the $k-$th atom's position.
For a ladder, a second set of position distributions $p^{(k),2}_{\rm pos}(\mathbf{r})$ would be added with the same Gaussian form up to $r_2 \to r_2 - R_0$.

\section{Distribution of the distances and interactions for a single chain.} 
Here we focus on a single one-dimensional chain as most of the properties which affect the results in the main text are due to the presence of an extended longitudinal direction. Still, the considerations made for the marginal distributions for pairs of atoms directly apply to any regular lattice configuration as well.
The distribution of the differences $\mathbf{d}_k = \mathbf{r}_{k+1} - \mathbf{r}_k = (d_{k,1}, d_{k,2}, d_{k,3})$ can be found in the Supplemental Material of Ref.~\cite{a_Marcuzzi_PRL_17} and, for isotropic traps, reads
\begin{widetext}
\be
\begin{split}
	p_{\rm diff}(\mathbf{d}_1 , \ldots , \mathbf{d}_{L-1}) = \int \lqq \prodl{k=1}{L} \rmd^3 r_k \, p^{(k)}_{\rm pos}(\mathbf{r}_k) \rqq  
	%\times \\ \times 
	\lqq  \prodl{k'=1}{L-1} \delta^{(3)} \lt  \mathbf{d}_{k'} - \lt \mathbf{r}_{k'+1} - \mathbf{r}_{k'} \rt  \rt \rqq = \\
	= \lqq  \frac{\sigma^{1-L}}{\sqrt{L} \lt \sqrt{2\pi } \rt^{L-1}}  \rqq^3    \rme{- \frac{1}{2\sigma^2} \sum_{k,q} \lqq   d_{k,1} A_{kq} d_{q,1}  +   d_{k,2} A_{kq} d_{q,2}  +  (d_{k,3} - R_0) A_{kq} (d_{q,3} - R_0)  \rqq } ,   
	\label{eq:pdiff}
\end{split}
\ee
\end{widetext}
where $A_{kq} = L - \max(k,q) - (L-k)(L-q)/L = (L - \max(k,q)) \min(k,q) / L$. The correlations between different components $d_{k,i}$ can be worked out via the inverse \cite{MatInverse}
\be
C = A^{-1} =   \matb{ccccc}    
	2 & -1 & 0 & 0 &   \\
	-1 & 2 & -1 & 0 &   \\
	0 & -1 & 2 & -1 &  \cdots  \\
	0 & 0 & -1 & 2 &    \\
	  &   &  \vdots &    & \ddots
\mate,
\label{eq:matC}
\ee
implying, 
\be
	\av{d_{k,i} d_{q,j}} - \av{d_{k,i}} \av{ d_{q,j}} = \sigma^2 \delta_{ij} \lt  2 \delta_{k,q} - \delta_{k,q+1} - \delta_{k,q-1}  \rt.
\ee
Subsequent distances are therefore (anti-)correlated, and these correlations pass onto any (non-trivial) function of the distances, and in particular the energy displacements $\delta V_k = V(d_k) - V(R_0)$.

As a consistency check, we remark that $C(L)$ is a $(L-1)\times (L-1)$ matrix, whose determinant satisfies the recursion relation
\be
	\det C(L) = 2\det C(L-1) - \det C(L-2) 
\ee
with ``seed'' (or initial conditions) $\det C(2) = 2$ and $\det C(3) = 3$, which is solved by $\det C(L) = L$. Consequently, the factor $\lt \sqrt{\det A} \rt^3$ produced by the Gaussian integration over all variables exactly cancels the $L^{-3/2}$ appearing in the normalization factor, as expected.

%All the marginal distributions for the single $\mathbf{d}_k$s, obtained via integration of the other $L-2$ variables from equation \eqref{eq:pdiff}, are equivalent and read
%\be
%	p_{\rm diff} (\mathbf{d}) = \frac{1}{(4\pi)^{3/2} \sigma_1 \sigma_2 \sigma_3 } \rme{-\frac{1}{4}  \lqq \frac{d_1^2}{\sigma_1^2}   +\frac{d_2^2}{\sigma_2^2} + \frac{(d_3 -r_0)^2}{\sigma_3^2}  \rqq         }.
%	\label{eq:distr_simpl}
%\ee
%(obtained, e.g., from \eqref{eq:pdiff} in App.\ref{app:A} by integrating over all but one variable). The effect on the Lyapunov exponent is shown in Fig.~\ref{fig:lyap} and this approximation seems to get reasonably close to the original result. For generic $\sigma_\perp \neq \sigma_\parallel$ it is not possible to obtain a closed expression for the distribution of the distance $d = \sqrt{d_x^2 + d_y^2 + d_z^2}$, which can be expressed however as the integral

\subsection{Marginal distribution for a single pair of atoms}

The $\mathbf{d}_k$s are identically distributed, so we can select any given one (and drop its index for brevity) and integrate over the other $L-2$ variables from equation \eqref{eq:pdiff}. This yields
\be
\begin{split}
	p_{\rm diff} (\mathbf{d}) &= \frac{1}{(4\pi)^{3/2} \sigma^3 } \rme{-\frac{1}{4 \sigma^2}  \lqq d_1^2   +d_2^2 + (d_3 - R_0)^2  \rqq         } = \\
	&=\frac{1}{(4\pi)^{3/2} \sigma^3 } \rme{-\frac{1}{4 \sigma^2}  \lqq d^2 + R_0^2 - 2 d_3 R_0  \rqq         },
	\label{eq:distr_simpl}
\end{split}
\ee
where $d = \abs{\mathbf{d}}$ denotes the distance between a pair of neighboring atoms. The distribution for this new variable can be then obtained via a solid angle integration and reads 
\be	
\begin{split}
	p_{\rm dist} (d) &=  \frac{d^2}{4(\pi)^{1/2} \sigma^3}  \rme{- \frac{1}{4\sigma^2} (d^2 + R_0^2) }  \int_0^\pi \rmd \theta   \sin\theta \,  \rme{-\frac{d R_0 \cos\theta }{2\sigma^2}}=   \\
	&= \frac{d}{\sqrt{\pi} \sigma R_0}  \rme{- \frac{1}{4\sigma^2} (d^2 + R_0^2) } \sinh \lt \frac{d R_0}{2\sigma^2} \rt.
\end{split}
\ee
The distribution of an energy shift $\delta V$ is now just a change of variables ($d \to d(\delta V)$) away, according to $P(\delta V) = \abs{d'(\delta V)} p_{\rm dist} (d(\delta V))$.
%%%
For the sake of generality, we keep $\alpha$ generic in 
\be
	d(\delta V) = \lt \frac{C_\alpha}{V_0 + \delta V} \rt^\frac{1}{\alpha},
\ee
where $V_0 = C_\alpha / R_0^\alpha$, which implies
\be
	d'(\delta V) = -\frac{1}{\alpha} \frac{C_\alpha^{1/\alpha}}{(V_0 + \delta V)^{1+1/\alpha}}.
\ee
Hence, the distribution of energy shifts for a pair is
\be
\begin{split}
	P (\delta V | V_0, R_0, \sigma) = \frac{\frac{R_0}{\sigma} }{\alpha \sqrt{\pi} V_0 \lt 1 + \frac{\delta V}{V_0}  \rt^{1 + \frac{2}{\alpha}}} \times \\
	 \times \ \rme{-\frac{R_0^2}{4\sigma^2}  \lqq   1 + \lt 1 + \frac{\delta V}{V_0}  \rt^{-\frac{2}{\alpha}} \rqq} \times \\
	 \times \ \sinh \lqq \frac{R_0^2}{2\sigma^2}  \lt 1 + \frac{\delta V}{V_0}  \rt^{-\frac{1}{\alpha}}  \rqq.
\end{split}
\ee
It is relatively simple to see that, if we define the dimensionless quantities $\delta v = \delta V / V_0$ and $s = \sigma/R_0$, we can simplify this expression further:
\be
\begin{split}
	P \lt \delta v | s \rt = \frac{1 }{\alpha \sqrt{\pi} s \lt 1 + \delta v  \rt^{1 + \frac{2}{\alpha}}} \times \\
	 \times \ \rme{-\frac{1}{4s^2}  \lqq   1 + \lt 1 + \delta v \rt^{-\frac{2}{\alpha}} \rqq} \times \\
	 \times \ \sinh \lqq \frac{1}{2s^2}  \lt 1 + \delta v  \rt^{-\frac{1}{\alpha}}  \rqq.
\label{eq:simpl}
\end{split}
\ee

The probability distribution function in Eq.~\eqref{eq:simpl} is defined in the domain \change{$\delta_v \in [-1,+\infty)$; for $\delta v = - 1 + \varepsilon$}, in the limit $\varepsilon \to 0^+$ it behaves as
\be 
	P\lt \delta v | s \rt \propto \varepsilon^{-1-\frac{2}{\alpha}} \rme{-\frac{1}{4s^2} \varepsilon^{-\frac{2}{\alpha}}} \sinh \lqq  \frac{\varepsilon^{-\frac{1}{\alpha}}}{2s^2} \rqq \to 0,
\ee
as the (vanishing) exponential factor dominates. In the opposite limit $\delta v \to \infty$, instead, the distribution behaves asymptotically as
\be
	P \lt \delta v | s \rt \approx \frac{1}{2\alpha \sqrt{\pi} s^3} \rme{-\frac{1}{4s^2}} \delta v^{-1 - 3/\alpha}.
	\label{eq:asympt2}
\ee
This shows that this distribution is fat-tailed. In particular, all the distribution moments $\av{\delta v^\beta}$ with $\beta \geq 3/\alpha$ are not defined and, for both $\alpha = 3$ (dipole-dipole interactions) and $\alpha = 6$ (van der Waals), this includes all integer moments (e.g., the mean and variance). These fat tails are the consequence of the approximation of an atom's position distribution as a Gaussian everywhere in space, i.e., including points much further away from the center of a trap than a few $\sigma$s. In other words, it appears to be an artifact of the description, rather than something \change{occurring in a real experiment}. The result of this approximation is to allow for an extremely small (but not vanishing) probability that two atoms can be arbitrarily close, which, due to the algebraic scaling of the interactions, produces considerable energy shifts. Moments like the mean and variance are therefore dominated by the very rare events in which two atoms lie very close to each other. The rarity of such events is encoded in the exponential suppression $\rme{-(1/4s^2)}$ in Eq.~\eqref{eq:asympt2}. In principle, these unphysical fat tails could affect our results, as it is known that, in the Anderson problem, the scaling of the localization length is modified when Cauchy-like distributions are chosen instead of more regular ones. However, as mentioned above, the fat tails in our case are strongly suppressed and one needs to assess how likely it is to actually probe them in a simulation or an experiment. For that purpose, let us first notice that the asymptotic behavior reported in \eqref{eq:asympt2} emerges when the argument of the $\sinh$ function in Eq.~\eqref{eq:simpl} is small, i.e., still assuming $\delta v \gg 1$, for
\be
	\delta v \gg \lt 2s^2 \rt^{-\alpha}.
\ee
Let us calculate now the probability of generating an energy shift within the tails, i.e.,
\be 
\begin{split}
	\mathbb{P}_s \equiv \mathbb{P} \lt \delta v > \lt 2s^2 \rt^{-\alpha} \rt = \int_{\lt 2s^2 \rt^{-\alpha}}^\infty \rmd \delta v \, P(\delta v | s).
\end{split}
\ee
Employing now the asymptotic expression \eqref{eq:asympt2} we obtain
\be
	\mathbb{P}_s = \mathbb{P} \lt \delta v > \lt 2s^2 \rt^{-\alpha} \rt  \approx \frac{4s^3}{3 \sqrt{\pi} } \rme{-\frac{1}{4s^2}} . 
\ee
This result apparently does not depend on $\alpha$, but we need to remember that the derivation assumes $\delta v \gg 1$, and is therefore only consistent if $(2s^2)^{-\alpha} \gg 1$. Considering $\alpha = 3$ or $6$, though, this is satisfied already for rather large disorder amplitudes, e.g., $s = 0.3$, which then yields $\mathbb{P}_{0.3} \approx 0.0013$. Due to the exponential factor, these probabilities decrease very fast with $s$. For $s = 0.1$, for instance, we get $\mathbb{P}_{0.1} \approx 10^{-14}$ and in the range spanned in the plots reported in the main text $s \leq 5 \times 10^{-4}$ this becomes $\mathbb{P}_{s} \ll 10^{-400000} $, which is clearly impossible to observe in any reasonable experiment or numerical procedure. Hence, we can safely assume the unphysical fat tails to be completely irrelevant in the determination of our numerical results in the regime considered.

\subsection{Distribution bias towards positive energy shifts}

As can be observed from Fig. 3 in the main text, there appears to be a bias of the distribution towards positive energy shifts \change{for small disorder}, which makes the features present in the localization length plots bend towards higher energies. Considering the marginal discussed in the previous section, this seems counter-intuitive, since the distribution \eqref{eq:simpl} seems to shift, for increasing $s$, towards negative values instead, eventually becoming peaked very close to $\delta v \approx -1$. It is also possible to provide an intuitive explanation for this, since, as two neighboring traps becomes wider and wider, it becomes much more likely for two atoms to lie at a larger distance than the one separating the two centers. The limiting case $s \gg R_0$ is indicative, as one can imagine that the atoms' positions can be picked uniformly in space on length scales $\gg R_0$.

Although we do not hold at the moment a convincing explanation of why the shift seems to point in the opposite direction, we believe it is due to the correlated nature of the full distribution (which indeed would be consistent with not seeing the correct behavior in a marginal) and we provide a physical argument which partially supports this conjecture. For simplicity, we shall work here with a chain, rather than a ladder. We hypothesize that (A) the number $L + 1$ of atoms is large, i.e., $L + 1 \gg 1$ (this choice is such that the number of distances is $L$); (B) the disorder is extremely small $s \lll 1$. A useful simplification from (B) is that we can effectively reduce the dimensionality of the problem and only consider the position displacement along the $z$ direction: in fact, if we write $\mathbf{d} = (\delta x, \delta y, R_0 + \delta z)$, then 
\be
	d = \abs{\mathbf{d}} = R_0 + \delta z + O(\delta x^2, \delta y^2, \delta z^2) \approx R_0 + \delta z.
\ee 
Hence, we can extract the probability of the distances $d_k$ directly from Eq.~\eqref{eq:pdiff}:
\be
\begin{split}
	p_{\rm dist} (d_1, \ldots , d_L) =   \frac{\sigma^{-L}}{\sqrt{L+1} \lt \sqrt{2\pi} \rt^{L}}   \times \\
	\times \rme{- \frac{1}{2\sigma^2} \sum_{k,q}  (d_{k} - R_0) A_{kq} (d_{q} - R_0)   },
\end{split}
\ee
with the same $A_{k,q}$ (up to increasing $L$ by $1$). Clearly, the marginal for any given variable $d_k$ is also Gaussian and its mean and variance can be straightforwardly extracted from the expression above:
\be
	\av{d_k} = R_0 \mand \sqrt{\av{d_k^2} - R_0^2} = \sqrt{\sigma^2 (A^{-1})_{kk}} = \sqrt{2} \sigma. 
\ee
These variables are therefore identically distributed and, due to the boundedness of the covariance (see matrix $C$ in Eq.~\eqref{eq:matC}), satisfy a generalized weak law of large numbers, as we demonstrate below for this very special case: let us define $D = (\sum_k d_k) / L$ and consider the probability $\mathbb{P} (\abs{D - R_0} > \varepsilon)$ of a fluctuation $\> \varepsilon$ around the mean value. By Chebyshev's inequality,
\be
	\mathbb{P} (\abs{D - R_0} > \varepsilon) \leq \frac{{\rm Var} D}{\varepsilon^2},
	\label{eq:Cheby}
\ee
where 
\be
\begin{split}
	{\rm Var} D & = \frac{\av{ \lt \sum_k \lt d_k - R_0 \rt \rt^2}}{L^2} = \\
	 &= \frac{1}{L^2} \sum_{k,q} \av{(d_k - R_0)(d_q - R_0)} =   \frac{\sigma^2}{L^2} \sum_{k,q} C_{k,q},
\end{split}
\ee
where $C_{k,q}$ are the elements of the matrix $C = A^{-1}$ in Eq.~\eqref{eq:matC}. This sum is not difficult to calculate, since each row but the first and the last totals $0$, whereas the first and last contribute $1$ each, implying ${\rm Var} D = 2 \sigma^2 / L^2$. As a consequence, by choosing a sufficiently large $L$ the r.h.s.~in Eq.~\eqref{eq:Cheby} can be made arbitrarily small or, more precisely, $\forall \delta >0 \,\, \exists \bar{L} : \forall L > \bar{L}$
\be 
	\mathbb{P}(\abs{D - R_0} > \varepsilon) \leq \delta,
\ee
and therefore $D \to R_0$ in probability. Note that, had we been dealing with independent variables, we would have retrieved a result where ${\rm Var} D \sim 1/L$ instead of $1/L^2$. On a less formal level, this can be understood as follows: consider that, in this effective one-dimensional picture, the sum of all distances corresponds to the distance between the first and last atoms. When generating the positions independently, this variable will not be affected by the random nature of the positions of all the intermediate ones; instead, if one were to generate the distances as independent variables, these would effectuate a random walk (with drift $R_0$) and thus the effective uncertainty in the position of the last atom, assuming knowledge of the first one, would be of order $O(\sqrt{L})$ and increase with the length of the chain.

Now, consider that, having chosen repulsive interactions ($V(R) > 0$), the interaction potential is a convex function. Hence, we can write down Jensen's inequality as
\be
	\frac{\sum_k V(d_k)}{L} \geq V\lt  \frac{\sum_k d_k}{L}  \rt.
\ee
By the weak law of large numbers, for large $L \gg 1$ we can effectively replace the r.h.s.~ of the inequality above with $V(R_0)$, which leaves us with the approximate statement
\be
	\sum_k V(d_k) \gtrsim L V(R_0),
\ee
i.e.,
\be
	\sum_k \lt V(d_k) - V(R_0) \rt \gtrsim 0,
\ee
i.e.,
\be
	\sum_k \delta V_k \gtrsim 0.
\ee
Albeit not a rigorous proof, this argument provides a clear indication that, for sufficiently large system sizes, the correlations among the variables will make positive biases in the energy shifts preferable to negative ones, in agreement with the qualitative features observed in the main text. Accepting this claim, there must be at least a point $s > 0$ where this bias is strictly positive. It then follows that there exists a right neighborhood of $s = 0$ in which the bias increases with $s$.

\section{Hilbert space reductions and restricted Hamiltonians}
The Hamiltonian introduced in the main text reads 
\begin{align}
 \op{H} = \underbrace{\Omega \, \sum_k^N  \op{\sigma}_x^{(k)} }_{\op{H}_1}\, + \, \underbrace{ \Delta\, \sum_k^N\,\op{n}_k +\,  \,
 \sum_{\substack{k= 1\\ m \ne k}}^N \, \ha V(d_{km}) \, \op{n}_m\, \op{n}_k }_{\op{H}_0},
 \label{Eq:Hamil_full}
\end{align}
where $d_{km}$ denotes the distance between the $k$-th and $m$-th atoms. In order to exploit the large energy separations present in the system, we switch to the interaction picture
\be
\begin{split}
	\op{H}_I (t) = \rme{i\op{H}_0 t} \op{H}_1 \rme{-\rmi\op{H}_0 t} = \Omega \sum_k \rme{\rmi\op{H}_0 t} \op{\sigma}^{(k)}_x \rme{-\rmi\op{H}_0 t}. 
\end{split}
	\label{eq:HI}
\ee
Recalling that $\comm{\op{\sigma}_x^{(k)}}{\op{n}_m} = 0$ for every $k \neq m$ and that $\op{\sigma}_x^{(k)} \op{n}_k = (1 - \op{n}_k) \op{\sigma}_x^{(k)}$ we can simplify the $k$-th addend in Eq.~\eqref{eq:HI}
\be
\begin{split}
	\rme{\rmi\op{H}_0 t} \op{\sigma}^{(k)}_x \rme{-\rmi\op{H}_0 t} & = \rme{\rmi t \op{n}_k (\Delta + \sum_{m \neq k} V(d_{km}) \op{n}_m)} \op{\sigma}_x^{(k)}  \times \\ 
	& \times \rme{-\rmi t \op{n}_k (\Delta + \sum_{m \neq k} V(d_{km}) \op{n}_m)} = \\
	& = \rme{\rmi t (2\op{n}_k - 1) (\Delta + \sum_{m \neq k} V(d_{km}) \op{n}_m)} \op{\sigma}_x^{(k)},
\end{split}
\ee
where in the first equality we singled out in the exponentials all the terms which depend upon $\op{n}_k$; all the remaining ones cancel out. The Hamiltonian $\op{H}_I$ can then be written as
\be
	\op{H}_I (t) = \Omega \sum_k \rme{\rmi t (2\op{n}_k - 1) (\Delta + \sum_{m \neq k} V(d_{km}) \op{n}_m)} \op{\sigma}_x^{(k)}.
\ee
\change{We apply a rotating-wave approximation} to discard all terms which oscillate fast in time. This implies that the oscillation frequency $\omega$ should be $\gg \Omega$ for a term to be neglected. Note that the frequency $\omega$ is however operator-valued:
\be
	\omega = (2\op{n}_k - 1) (\Delta + \sum_{m \neq k} V(d_{km}) \op{n}_m).
\ee
Since the prefactor $-1 \leq 2\op{n}_k - 1 \leq 1$ is of order $O(1)$, it is the second factor which is decisive for the selection. We introduce now for every site $k$ a projector $\op{P}_k$ over all states where there is a single excitation among the neighbors of $k$ and no additional one within a radius $2 R_0$. Its specific structure depends clearly on the structure of the lattice, but if we define by $\mal{F}_k$ the set of nearest-neighboring sites of $k$ and by $\mal{S}_k$ the set of sites within a distance $2R_0$ from $k$ which are neither site $k$ itself nor one of the sites in $\mal{F}_k$, then we can give an implicit definition according to
\be
	\op{P}_k = \sum_{q \in \mal{F}_k} \op{n}_q \prod_{q' \in \mal{F}_k, \\ q' \neq q} (1 - \op{n}_{q'}) \prod_{q'' \in \mal{S}_k} (1 - \op{n}_{q''}).
	\label{eq:defP}
\ee
Checking that the expression above satisfies $\lt \op{P}_k \rt^2 = \op{P}_k$ is straightforward if one recalls that $\op{n}_q^2 = \op{n}_q$ and $(1 - \op{n}_q)^2 = 1 - \op{n}_q$ $\forall \,\,q$. The relevance of the projector $\op{P}_k$ is that it precisely identifies the constraints -- identified in the main text -- under which a spin (or atom) is able to flip (or being excited/de-excited). Slightly more formally,
\be
	(\Delta + \sum_{m \neq k} V(d_{km}) \op{n}_m) \op{P}_k \approx (\Delta  - V(R_0)) \op{P}_k = 0,
\ee
where we have neglected all contributions from excitations beyond a distance of $2R_0$.
Furthermore, note that according to definition \eqref{eq:defP} $\op{P}_k$ acts trivially on site $k$ and thus commutes with all local operators which instead exclusively act on that site; in particular, $\comm{\op{\sigma}_x^{(k)}}{\op{P}_k} = 0$. Defining for brevity $\op{Q}_k = \mathbb{1} - \op{P}_k$ the projector onto the orthogonal subspace ($\op{Q}_k^2 = \op{Q}_k$, $\op{Q}_k \op{P}_k = 0$) we thus have
\be
\begin{split}
	 \op{\sigma}_x^{(k)} &= \lt \op{P}_k + \op{Q}_k \rt	\op{\sigma}_x^{(k)}\lt \op{P}_k + \op{Q}_k \rt = \\
	&= \op{P}_k \op{\sigma}_x^{(k)}\op{P}_k + \underbrace{\op{Q}_k \op{\sigma}_x^{(k)}\op{P}_k}_{=0} + \underbrace{\op{Q}_k \op{\sigma}_x^{(k)}\op{P}_k}_{=0} + \op{Q}_k \op{\sigma}_x^{(k)}\op{Q}_k = \\
	& = \op{P}_k \op{\sigma}_x^{(k)} + \op{Q}_k \op{\sigma}_x^{(k)}.
\end{split}
\ee
Hence, we can separate the interaction Hamiltonian $\op{H}_I$ into two contributions:
\be
\begin{split}
	\op{H}_I(t) &\approx \Omega \sum_{k} \op{P}_k \op{\sigma}_x^{(k)} + \\
	&+ \rme{\rmi t (2\op{n}_k - 1) (\Delta + \sum_{m \neq k} V(d_{km}) \op{n}_m)} \op{Q}_k \op{\sigma}_x^{(k)}.
\end{split}
\ee
The space of configurations onto which $\op{Q}_k$ has support can be further split into three classes:
\begin{itemize}
	\item[(A)] States where site $k$ has two or more excited nearest neighbors;
	\item[(B)] States where site $k$ has only one excited neighbor, but there is at least another excitation within a radius $2R_0$;
	\item[(C)] States where no neighbors of $k$ are excited.
\end{itemize}
In case (A) the interaction potential on site $k$ is $\geq 2V(R_0)$; accounting for the \change{facilitation condition} $\Delta = -V(R_0)$ we find $\omega \gtrsim V(R_0) \gg \Omega$; these terms are thereby oscillating very fast and can be discarded. Terms of type (B) are facilitated by the single neighboring excitation, but the presence of an additional one within a distance $2R_0$ implies that
\be
	\Delta + \sum_{m \neq k} V(d_{km}) \op{n}_m \geq V(2R_0) 
\ee
and therefore $\omega \gtrsim V(2R_0) \gg \Omega$, which allows us to neglect all type-(B) contributions as well. Terms belonging to class (C) are instead more delicate, since an appropriate combination of the interactions with many excitations at different distances could approximately cancel out the detuning $\Delta$. For instance, for dipole-dipole interactions ($\alpha = 3$) the potential obeys $V(\gamma R_0) = V(R_0) \gamma^{-3}$; considering a honeycomb lattice with $5$ excited next-nearest neighbors at distance $R_1 = \sqrt{3} R_0$ and a single excited next-next-next-next-nearest (or fourth-nearest for brevity) neighbor at distance $R_4 = 3R_0$ one finds 
\be
\begin{split}
	\Delta + \sum_{m \notin \mal{F}_k} V(d_{km}) \to - V(R_0) + 5 V(R_1) + V(R_4) = \\
	=  V(R_0) \lt  -1 + \frac{5}{3\sqrt{3}} + \frac{1}{3^3} \rt \approx -0.00071 \, V(R_0).
\end{split}
\ee
However, configurations such as the one described above always require a large local density of excitations, and hence can only affect Hilbert subspaces at higher energies than the ones considered in the main text, separated at least by some factors of $V(R_1) \gg \Omega$. As long as we consider the low-energy Hilbert subspaces, it is thus fine to neglect terms of type (C) as well. Overall, in the subspaces we are \change{interested in we can approximate}
\be 
	\op{H}_I (t) \approx \Omega \sum_k \op{P}_k \op{\sigma}_x^{(k)}.
\ee
Going back to the original Schr\"odinger representation is now straightforward and yields
\be
	\op{H} \approx \Omega \sum_k \op{P}_k \op{\sigma}_x^{(k)} + \Delta \sum_k \op{n}_k +  \sum_{\substack{k= 1\\ m \ne k}}^N \, \ha V(d_{km}) \, \op{n}_m\, \op{n}_k.
\ee
Note that in the specific subspace (let us call it $\mal{H}_1$) considered in the main text, the one including all possible one-excitation states plus all possible pairs of neighboring ones, the diagonal part $\op{H}_0$ acts trivially as the null operator and can thus be discarded, implying
\be
	\op{H}_{\mal{H}_1} = \Omega \sum_k \op{P}_k \op{\sigma}_x^{(k)}.
\label{eq:HH1}
\ee
We remark that the same derivation can be followed in the presence of weak disorder by changing the definition of $\op{H}_1$ in Eq.~\eqref{Eq:Hamil_full} to
\be
	\op{H}_1 = \Omega \sum_k  \op{\sigma}_x^{(k)} + \ha \sum_{k\neq q} \delta V (d_{kq}) \op{n}_k \op{n}_q.
\ee
Since the second term is diagonal and commutes with $\op{H}_0$, the calculation of the interaction picture is straightforward:
\be
\begin{split}
	\op{H}_I (t) &= \Omega \sum_k \rme{\rmi t (2\op{n}_k - 1) (\Delta + \sum_{m \neq k} V(d_{km}) \op{n}_m)} \op{\sigma}_x^{(k)} + \\
	&+ \ha \sum_{k\neq q} \delta V (d_{kq}) \op{n}_k \op{n}_q
\end{split}
\ee
and one can follow the same steps outlined above.

\subsection{Hilbert space lattice structure}

Having derived the restricted Hamiltonian \eqref{eq:HH1} we can now identify the geometric structure of the Hilbert space in the \change{basis of eigenstates of $\op{\sigma}_z^{(k)}$}. To start with, we introduce the following definitions for the basis itself: we call $\ket{M_k}$ states with a single excitation present on site $k$, whereas we denote by $\ket{N_{kq}}$ states with a pair of excitations on sites $k$ and $q$. Fixing the number $N$ of tweezers, the Hilbert subspace we work in is therefore defined as
\be
	 \mal{H}_1 = \rm{Span} \left\{  \ket{M_k}, \ket{N_{kq}} |\,  k = 1 , \ldots , N ; q \in \mal{F}_k  \right\},
\ee
where we recall that $\mal{F}_k$ is the set of nearest neighbors of site $k$. Note that, since $\ket{N_{kq}} = \ket{N_{qk}}$ the pair states are doubly counted; however, this clearly still leads to the generation of the same vector space. Alternatively, one can also define an equivalence relation $\ket{N_{kq}} \sim \ket{N_{ml}} \Leftrightarrow (k=m \wedge q = l)\vee(k=l \wedge q = m)$ and take the quotient of the r.h.s.~above. In the following, it is understood that the states $\ket{N_{kq}}$ are always taken from this space, i.e., we shall never consider states with two isolated excitations at distance $d > R_0$. 

By construction, $\op{H}_{\mal{H}_1} \mal{H}_1 \subseteq \mal{H}_1$. Furthermore, we know that the action of $\op{P}_k \op{\sigma}_x^{(k)}$ is to flip the spin in site $k$ conditioned on the presence of a single excitation in $\mal{F}_k$ and no additional one in $\mal{S}_k$. This implies 
\be
	\op{P}_k \op{\sigma}_x^{(k)} \ket{M_l} = \sysb{lcc} 0 & \text{if} & l = k,  \\
													\ket{N_{kq}} & \text{if} & l \in \mal{F}_k, \\													0 & \text{otherwise}. & \  \syse 
\ee
Considering that $l \in \mal{F}_k \Leftrightarrow k \in \mal{F}_l$, one can see that
\be
	\op{H}_{\mal{H}_1} \ket{M_l} = \Omega \sum_{k \in \mal{F}_l} \ket{N_{kl}}. 
\ee

Similarly, 
\be
	\op{P}_k \op{\sigma}_x^{(k)} \ket{N_{ql}} = \sysb{lcc} \ket{M_l} & \text{if} & q = k,  \\
													\ket{M_q} & \text{if} & l=k, \\													0 & \text{otherwise}, & \  \syse 
\ee
since by construction the only facilitated spins are in sites $q$ and $l$. Hence,
\be
	\op{H}_{\mal{H}_1} \ket{N_{ql}} = \Omega \lt \ket{M_q} + \ket{M_l} \rt.
\ee

Collecting these considerations, we can find the Hamiltonian matrix elements:
\begin{subequations}
\begin{align}
	\bra{M_q} \op{H}_{\mal{H}_1} \ket{M_k} &= 0 \label{eq:mat_elem1}\\
	\bra{N_{ml}} \op{H}_{\mal{H}_1} \ket{N_{kq}} & = 0 \\
	\bra{N_{ml}} \op{H}_{\mal{H}_1} \ket{M_k} &= \sysb{lcc} \Omega & \text{if} & l = k,  \\
													\Omega & \text{if} & m=k, \\													0 & \text{otherwise}. & \  \syse 
													\label{eq:mat_elem3}
\end{align}
\end{subequations}
Now, there are as many states $\ket{M_k}$ as there are sites, so it is natural to make a connection: starting from the \change{real-space} geometry of the tweezer array, which defines the original lattice structure, we place for visual aid each state $\ket{M_k}$ on the corresponding site $k$. Crucially, each pair state $\ket{N_{kq}}$ is exclusively connected (via the Hamiltonian) to the two one-excitation states $\ket{M_k}$ and $\ket{M_q}$, so it is \change{placed as a mid-point} between sites $k$ and $q$, changing the structure to a generalized Lieb lattice. Now, by drawing a link between any pair of sites every time the corresponding states yield a non-zero Hamiltonian matrix element one precisely reconstructs the kind of lattices we displayed in Fig.~$1$ in the main text.

\section{Bound on the number of flat bands}

\change{We provide an account of the lower bound of the number of flat bands $n_{\rm flat} \geq \abs{n_1 - n_2}$ mentioned in the main text.} We recall that $n_{\rm flat}$ denotes the number of flat bands in the model, $n_1$ the number of one-particle states per unit cell and $n_2$ the corresponding number of pair states per unit cell. Before doing that, however, we briefly comment on the fact that the spectrum of the hopping Hamiltonians \eqref{eq:HH1} is always symmetric with respect to $\epsilon=0$. In fact, one can define the parity transformation 
\be
	\op{U} = \op{U}^\dag = \lt -1 \rt^{\sum_k \op{n}_k}
\ee
which, in the subspace $\mal{H}_1$, acts according to $\op{U} \ket{M_k} = - \ket{M_k}$ on all one-excitation states and $\op{U} \ket{N_{kq}} = \ket{N_{kq}}$ on all pair states. Combined with Eqs.~\eqref{eq:mat_elem1}-\eqref{eq:mat_elem3}, this implies $\op{U}^\dag \op{H}_{\mal{H}_1} \op{U} = - \op{H}_{\mal{H}_1}$. Hence, if $\ket{\epsilon}$ is an eigenvector of the Hamiltonian at energy $\epsilon$, then $\op{U} \ket{\epsilon}$ is also an eigenvector, but at eigenvalue $-\epsilon$, proving the symmetry of the spectrum under reflection $\epsilon \to -\epsilon$.

We start directly from the synthetic lattice reconstructed in the Hilbert space according to the procedure described in the previous section. This structure is not in general a Bravais lattice and needs, as a first step, to be reduced to one by identifying an appropriate ``basis''. This is a standard procedure in crystallography and solid state physics and we refer the reader to any good introductory textbook (see e.g., \cite{Grosso}). For the reader's convenience, we however recall here just a few of the most basic concepts: a Bravais lattice is a lattice structure where the positions $\vec{l}$ of the lattice sites can be written as discrete combinations
\be
	\vec{l} = \sum_{i=1}^d z_i \vec{a}_i  \ \ \ \text{with} \ \ z_i \in \Z.
	\label{eq:realvec}
\ee
 of a set of $d$ linearly-independent \emph{primitive lattice vectors} $\vec{a}_i$ ($i = 1 \ldots d$), where $d$ is the dimensionality of the system. If a site is located at the origin, all sites can be found this way and all points at positions $\vec{l}$ are lattice sites. Any lattice is, by definition, a periodically repeating pattern, and is therefore invariant under a certain set of translations by $\vec{l}$ for some specific choice of the primitive lattice vectors. However, in many cases an additional set of $B$ vectors $\set{\vec{b}_1 , \ldots \vec{b}_B}$, called ``basis'', is required. In such cases, and fixing conventionally $\vec{b}_1 = 0$ which can be done without loss of generality, if one lattice point is located at the origin, every point at a position $\vec{l}$ is also a lattice site, but not all lattice sites are at positions $\vec{l}$. All of them are instead found at positions $\vec{l} + \vec{b}_j$ with $j =1, \ldots ,B$. We also remark that distances between sites in the synthetic lattice are not meaningful, being just a convenient way to visualize the structure of the Hilbert space. Hence, we are free to rescale the length of all (dimensionless) vectors $\vec{a}_i$, $\vec{b}_j$ by a common factor. In all the examples discussed below the primitive lattice vectors have the same length and we shall choose to normalize them to unit length ($\abs{\vec{a}_i} = 1$). Also, for brevity in the following we refer to the $\R^d$ space where these vectors live as the \emph{direct space}.

We also introduce the reciprocal lattice vectors $\vec{a}_i^\ast$, $i = 1 \ldots d$ which satisfy the defining relations
\be
	\vec{a}_i^\ast \cdot \vec{a}_j = 2\pi \delta_{ij}.
\ee
The reciprocal Bravais lattice is then reconstructed by taking integer combinations of these vectors, i.e.,
\be
	\vec{G} = \sum_{i=1}^d z_i^\ast \vec{a}_i^\ast \ \ \ \text{with} \ \ z_i^\ast \in \mathbb Z.
	\label{eq:rec}
\ee
\change{We define a unit cell $\mal{U}^\ast$ which contains only one reciprocal lattice point. All possible translations $\vec{G}$ of  $\mal{U}^\ast$ cover the whole space $\R^d$ without any overlaps.
It can be visualized as a tessellation with $\mal{U}^\ast$ a tile.}
From a slightly different (but equivalent) perspective, one can define the equivalence relation between vectors $\vec{k}$, $\vec{q} \in \R^d$ living in reciprocal space
\be 
	\vec{k} \sim \vec{q} \Leftrightarrow \exists\, \vec{G} \,|\, \vec{k} = \vec{q} + \vec{G}
\ee
with $\vec{G}$ a \emph{reciprocal lattice} vector. Hence, the unit cell may be defined as a quotient $\R^d / \sim$. By defining quasi-momenta $\vec{k}$ as reciprocal space vectors belonging to a unit cell $\mal{U}^\ast$, one can define a Fourier series in the usual way for any generic quantity $A_{\vec{l}}$ living on the direct-space Bravais lattice  
\be
	\wt{A}_{\vec{k}} =  \sum_{\vec{l}} \rme{-\rmi \vec{k} \cdot \vec{l}} A_{\vec{l}} \,.
\ee
The corresponding inverse transform is also standard:
\be
	A_{\vec{l}} = \int_{\mal{U}^\ast} \frac{\rmd^d k}{(2\pi)^d} \,\rme{\rmi \vec{k} \cdot \vec{l}}\, \wt{A}_{\vec{k}} \,,
\ee
as can be shown remembering that
\be
	\frac{\vec{G} \cdot \vec{l}}{2\pi} \in \Z
\ee
and using the Poisson-summation-derived distributional identity
\be
	\sum_{z\in \Z} \rme{-\rmi \alpha z} = \sum_{m \in \Z} 2 \pi \delta (\alpha + 2\pi m),
\ee
with $\alpha \in \R$ and $\delta$ the Dirac delta. The choice of the unit cell is not unique; in the following we assume to be working in the \emph{first Brillouin zone} $\mal{B}$ \cite{Grosso}.

Clearly, the definitions above do not hinge upon working in a specific space and, indeed, one can analogously define a unit cell in direct space which contains a single Bravais lattice point. Hence, such a unit cell includes $B$ synthetic lattice points. It is quite natural to subdivide them according to whether they are of the ``one-excitation'' or ``pair'' kind. As done in the main text, we define $n_1$ the number of one-excitation states in a unit cell and $n_2 = B - n_1$ the number of pair ones. For example,
\begin{itemize}
	\item Synthetic square lattice (Lieb lattice): $n_1 = 1$, $n_2 = 2$, $B = 3$.
	\item Synthetic triangular lattice: $n_1 = 1$, $n_2 = 3$, $B = 4$.
	\item Synthetic honeycomb lattice: $n_1 = 2$, $n_2 = 3$, $B = 5$.
\end{itemize}
Since each synthetic lattice point can be uniquely associated to a given primitive lattice vector $\vec{l}$ and basis vector $\vec{b}_i$, we can unambiguously denote each state in the Hilbert subspace $\mal{H}_1$ as a tensor product $\ket{\vec{l}} \otimes \ket{\vec{b}_i}$. \change{For later convenience, we introduce now a new notation distinguishing between the basis vectors} identifying one-excitation states $\left(\ket{\vec{b}_i} \to \ket{\mu_j}\,,j=1 ,\ldots, n_1\right)$ and pair states
$ \left(\ket{\vec{b}_i} \to \ket{\nu_j}\,,j=1 ,\ldots, n_2\right)$, so that the space of basis states is equivalently generated as
\be
	Span \set{  \ket{\mu_1}, \ldots \ket{\mu_{n_1}}, \ket{\nu_1} , \ldots, \ket{\nu_{n_2}} }.
\ee
Consequently, there is a bijective correspondence between states $\ket{M_k}$ and states $\ket{\vec{l}} \otimes \ket{\mu_i}$ and between states $\ket{N_{kq}}$ and states $\ket{\vec{l}} \otimes \ket{\nu_i}$.

We also define the lattice translation operator $T_{\vec{j}}$, where $\vec{j}$ is a Bravais lattice vector, which acts on the positional degrees of freedom according to
\be
	T_{\vec{j}} \ket{\vec{l}} = \ket{\vec{l} + \vec{j}}.
\ee
By the straightforward quasi-momentum states definition
\be
	\ket{\vec{k}} =  \sum_{\vec{l}} \rme{-i \vec{k} \cdot \vec{l}}  \ket{\vec{l}}
\ee
one also gets
\be
	T_{\vec{j}} \ket{k} = \rme{\rmi\vec{k} \cdot \vec{j}} \ket{\vec{k}}.
\ee
The Hamiltonian can now be generically characterized as a sum of terms 
\be
\begin{split}
	\op{H}_{\mal{H}_1} = \Omega \sum_{\vec{l}}   \sum_{\vec{j}} \sum_{m= 1}^{n_1} \sum_{n=1}^{n_2} \lt C_{\vec{j},m,n}   \ket{\mu_m} \bra{\nu_n}   + \right. \\
	+ \left. D_{\vec{j},m,n}   \ket{\nu_n} \bra{\mu_m}   \rt \ket{\vec{l} + \vec{j}}  \bra{\vec{l}}  \, ,
\end{split}
\ee
where $C_{\vec{j}}$ and $D_{\vec{j}}$ are collections of connectivity matrices with elements $1$ (if two states are linked) and $0$ (if the two states are not). For instance, if the Hamiltonian can cause a hop from $\vec{l}$ to $\vec{l} + \vec{a}_1$ accompanied by a change $\ket{\mu_1} \to \ket{\nu_1}$, then $D_{\vec{a}_1,1,1} = 1$. Note that these are, in general, rectangular matrices of size $n_1 \times n_2$. Furthermore, to ensure that $H$ is hermitian they must satisfy
\be
	C_{-\vec{j},m,n} = D_{\vec{j},m,n}^\ast = D_{\vec{j},m,n},
\ee
where the last equality comes from the fact that they are defined to be real (their elements being either $0$ or $1$). Note that no terms $\propto \ket{\mu_m} \bra{\mu_n} $ or $\ket{\nu_m} \bra{\nu_n} $ appear, as one-excitation states are exclusively connected to pair ones and vice versa (see Eqs.~\eqref{eq:mat_elem1}-\eqref{eq:mat_elem3}).
\change{Neither $C$ nor $D$ depends explicitly on $\vec{l}$, as the form of the Hamiltonian is independent of the choice of the origin.} In this form, it is not difficult to exploit this  symmetry of the Hamiltonian under discrete lattice translations to partially diagonalize it in terms of Fourier modes:
\begin{widetext}
\be
\begin{split}
	\op{H}_{\mal{H}_1} & = \Omega \sum_{\vec{l}}   \sum_{\vec{j}} \sum_{m= 1}^{n_1} \sum_{n=1}^{n_2} \lt C_{\vec{j},m,n}   \ket{\mu_m} \bra{\nu_n}   + C_{-\vec{j},m,n}   \ket{\nu_n} \bra{\mu_m}   \rt T_{\vec{j}} \ket{\vec{l} }  \bra{\vec{l}} = \\
	& = \Omega    \sum_{\vec{j}} \sum_{m= 1}^{n_1} \sum_{n=1}^{n_2} \lt C_{\vec{j},m,n}   \ket{\mu_m} \bra{\nu_n}   + C_{-\vec{j},m,n}   \ket{\nu_n} \bra{\mu_m}   \rt T_{\vec{j}}  \sum_{\vec{l}} \ket{\vec{l} }  \bra{\vec{l}} = \\
	& = \Omega    \sum_{\vec{j}} \sum_{m= 1}^{n_1} \sum_{n=1}^{n_2} \lt C_{\vec{j},m,n}   \ket{\mu_m} \bra{\nu_n}   + C_{-\vec{j},m,n}   \ket{\nu_n} \bra{\mu_m}  \rt T_{\vec{j}}  \int_{\mal{B}}   \frac{\rmd^d k}{(2\pi)^d} \, \proj{\vec{k}}   = \\
	& = \Omega  \int_{\mal{B}}   \frac{\rmd^d k}{(2\pi)^d} \,  \sum_{\vec{j}} \sum_{m= 1}^{n_1} \sum_{n=1}^{n_2} \lt C_{\vec{j},m,n}   \ket{\mu_m} \bra{\nu_n}   + C_{-\vec{j},m,n}   \ket{\nu_n} \bra{\mu_m}   \rt \rme{\rmi\vec{k} \cdot \vec{j}}  \ket{\vec{k} }  \bra{\vec{k}} = \\
	& =  \Omega  \int_{\mal{B}}   \frac{\rmd^d k}{(2\pi)^d} \, \sum_{m= 1}^{n_1} \sum_{n=1}^{n_2} \lqq  \lt \sum_{\vec{j}} C_{\vec{j},m,n} \rme{\rmi\vec{k} \cdot \vec{j}}  \rt \ket{\mu_m} \bra{\nu_n} + \lt \sum_{\vec{j}} C_{\vec{j},m,n} \rme{\rmi\vec{k} \cdot \vec{j}}  \rt^\ast \ket{\nu_n} \bra{\mu_m}   \rqq \proj{\vec{k}} = \\
	& = \Omega \int_{\mal{B}}   \frac{\rmd^d k}{(2\pi)^d} \, \sum_{m= 1}^{n_1} \sum_{n=1}^{n_2} \lqq  \wt{C}_{-\vec{k},m,n}  \ket{\mu_m} \bra{\nu_n} + \lt \wt{C}_{-\vec{k},m,n} \rt^\ast \ket{\nu_n} \bra{\mu_m}   \rqq \proj{\vec{k}} ,
\end{split}
\ee
\end{widetext}
where again
\be
	\wt{C}_{-\vec{k},m,n} = \lt \sum_{\vec{j}} C_{\vec{j},m,n} \rme{\rmi\vec{k} \cdot \vec{j}}  \rt
\ee
is, for every $\vec{k} \in \mal{B}$, a rectangular $n_1 \times n_2$ matrix. Calling now
\be
	\op{M}_{\vec{k}} = \sum_{m= 1}^{n_1} \sum_{n=1}^{n_2} \lqq  \wt{C}_{-\vec{k},m,n}  \ket{\mu_m} \bra{\nu_n} + h.c.   \rqq,
\ee
we can represent it as a matrix in the basis $\set{\ket{\mu_1}, \ldots, \ket{\mu_{n_1}}, \ket{\nu_1}, \ldots, \ket{\nu_{n_2}}} $, which yields
\be
	M_{\vec{k}} = \matb{c|c}  0 & \wt{C}_{-\vec{k}}  \\[1mm] \hline \\[-3mm] \wt{C}^\dag_{-\vec{k}} & 0   \mate.
\ee
Due to this particular block structure,
\be
	{\rm{Rank}} \set{M_{\vec{k}}} = \rm{Rank} \set{\wt{C}_{-\vec{k}}} + \rm{Rank} \set{\wt{C}^\dag_{-\vec{k}}} .
\ee
Furthermore, the rank of a rectangular matrix is never greater than its shortest side. In this case,
\be
	{\rm{Rank}} \set{\wt{C}_{-\vec{k}}} \leq  \min \set{n_1, n_2},
\ee
which in turn implies that the rank of the square matrix $M_{\vec{k}}$ is $\leq 2 \min \set{n_1, n_2}$. This means that the size of the kernel of $M_{\vec{k}}$ has a lower bound
\be 
\begin{split}
	\dim  & \lt {\rm{Ker}} \, M_{\vec{k}} \rt  = B - {\rm{Rank}} \set{M_{\vec{k}}} \geq \\
	& \geq  \lt n_1 + n_2 \rt - 2 \min \set{n_1, n_2} =  \\
	& =  \max \set{n_1, n_2} -  \min \set{n_1, n_2} = \\
	& =  \abs{n_1 - n_2}.
\end{split}
\ee
Hence, if $\abs{n_1 - n_2} \geq 1$ then for every $\vec{k}$ one can find a kernel vector $\ket{v_{\vec{k}}}$ in the basis such that $\op{M}_{\vec{k}} \ket{v_{\vec{k}}} = 0$. Correspondingly, $\op{H}_{\mal{H}_1} \ket{\vec{k}} \otimes \ket{v_{\vec{k}}} = 0$ $\forall \vec{k}$ and the set of all these states forms a zero-energy flat band. Clearly, if $\abs{n_1 - n_2} > 1$ then more than one choice of $\ket{v}_{\ket{k}}$ can be made per each quasi-momentum $\vec{k}$, each identifying an independent flat band. Hence, calling the number of flat bands in the model $n_{\rm flat}$, consistently with the main text notation,
\be
	n_{\rm flat} = \dim   \lt {\rm{Ker}} \, M_{\vec{k}} \rt \geq \abs{n_1 - n_2},
\ee
which proves the bound.

The general rules for filling the matrix elements of $\wt{C}_{\vec{k}}$ are the following:
\begin{itemize}
	\item Choose n-th column $1 \leq n \leq n_2$.
	\item Consider the \emph{two} possible ways in which a particle can hop from the intermediate state $\ket{\nu_n}$ within the basis to its neighbors $\ket{\mu_m}$ and $\ket{\mu_p}$.
	\item Add $\rme{\rmi\vec{k}\cdot \vec{j}_n}$ to $C_{-\vec{k},m,n}$ and $\rme{\rmi\vec{k}\cdot \vec{j}_p}$ to $C_{\vec{k},p,n}$, where $\vec{j}_{m/p}$ are the lattice vectors pointing to the \emph{arrival} lattice sites.
\end{itemize}
In the next sections we work out some examples among the ones displayed in the main text. For simplicity, we set $\Omega = 1$.

%%%%%%%%%%%%%%%%%%%%%%%%%%%%%%%%%%%%%%%%%%%%%%%%%%%%%
%%%%%%%%%%%%%%%%%%%%%%%%%%%%%%%%%%%%%%%%%%%%%%%%%%%%%
%%%%%%%%%%%%%%%%%%%%%%%%%%%%%%%%%%%%%%%%%%%%%%%%%%%%%
%%%%%%%%%%%%%%%%%%%%%%%%%%%%%%%%%%%%%%%%%%%%%%%%%%%%%
%%%%%%%%%%%%%%%%%%%%%%%%%%%%%%%%%%%%%%%%%%%%%%%%%%%%%

\subsection{Example: the triangular lattice}

The triangular lattice is a two-dimensional Bravais lattice with primitive lattice vectors
\be
	\vec{a}_1 = a \lt 1,0   \rt^\intercal \mand \vec{a}_2 = a \lt \cos \frac{\pi}{3}, \sin \frac{\pi}{3}   \rt^\intercal ,
\ee
with $a$ the real-space lattice spacing. In the Hilbert space, we have again a triangular structure where a new site is added on each link. 
%In a periodic (or infinitely-extended) lattice there are $6$ bonds departing from every site and each connects two. Hence, if $N$ is the total number of sites then the lattice has $3N$ bonds.

It is not difficult to see that this reduces to a pure triangular lattice by choosing a basis of $4$ sites, a single one-excitation one ($n_1 = 1$) and $3$ pair ones ($n_2 = 3$). The primitive lattice vectors will be the same as above, where we fix for simplicity $a=1$. The basis states can be chosen according to:
\begin{itemize}
	\item[$\ket{\mu_1}$]: a one-excitation site at $\vec{b} = 0$.
	\item[$\ket{\nu_1}$]: a pair site at $\vec{b} = \vec{a}_1 / 2$.
	\item[$\ket{\nu_2}$]: a pair site at $\vec{b} = \vec{a}_2 / 2$.
	\item[$\ket{\nu_3}$]: a pair site at $\vec{b} = (\vec{a}_1 - \vec{a}_2) / 2$.
\end{itemize}
The matrix $\wt{C}_{-\vec{k}}$ is now a $1 \times 3$ matrix whose elements can be computed via the procedure outlined above:
\begin{itemize}
	\item[$\wt{C}_{-\vec{k},1,1}$]: from basis state $\ket{\nu_1}$ one can reach state $\ket{\mu_1}$ within the same Bravais lattice site ($\Rightarrow +1$) or state $\ket{\mu_1}$ at the neighboring site $\vec{j} = \vec{a}_1$ ($\Rightarrow +\rme{\rmi\vec{k} \cdot \vec{a}_1}$).
	\item[$\wt{C}_{-\vec{k},1,2}$]: from basis state $\ket{\nu_2}$ one can reach state $\ket{\mu_1}$ within the same site ($\Rightarrow +1$) or state $\ket{\mu_1}$ at the neighboring site $\vec{j} = \vec{a}_2$ ($\Rightarrow +\rme{\rmi\vec{k} \cdot \vec{a}_2}$).
	\item[$\wt{C}_{-\vec{k},1,3}$]: from state $\ket{\nu_3}$ one can reach state $\ket{\mu_1}$ within the same site ($\Rightarrow +1$) or state $\ket{\mu_1}$ at the neighboring site $\vec{j} = \vec{a}_1 - \vec{a}_2$ ($\Rightarrow +\rme{\rmi\vec{k} \cdot (\vec{a}_1 - \vec{a}_2)}$).
\end{itemize}
Collecting all terms, the matrix $\wt{C}_{-\vec{k}}$ reads
\be
	\wt{C}_{-\vec{k}} = \matb{ccc} 1 + \rme{\rmi\vec{k} \cdot \vec{a}_1}, &   1 + \rme{\rmi\vec{k} \cdot \vec{a}_2} , &  1 + \rme{\rmi\vec{k} \cdot( \vec{a}_1 - \vec{a}_2)}   \mate \equiv \vec{w}_{\vec{k}}^\dag
\ee
and is equivalent to a three-dimensional vector $\vec{w}_{\vec{k}}$. Thus, the total matrix $M_{\vec{k}}$ can be expressed as 
\be
	M_{\vec{k}} = \matb{c|c} 0 & \vec{w}_{\vec{k}}^\dag  \\ \hline \\[-2mm] \vec{w}_{\vec{k}} & 0    \mate.
\ee
There are two kernel states corresponding to four-dimensional vectors $(0,v_{\vec{k},1})$ and $(0,v_{\vec{k},2})$ with $\vec{w}^\dag_{\vec{k}} \cdot \vec{v}_{\vec{k},1/2} = 0$. These states thus reconstruct two flat bands, in line with the bound $n_{\rm flat} \geq 2$ of this case.
 
The remaining two bands can be calculated instead by squaring ${M}_{\vec{k}}$:
\be
	M_{\vec{k}}^2 =  \matb{c|c} \vec{w}^\dag_{\vec{k}} \cdot \vec{w}_{\vec{k}} & 0  \\ \hline \\[-2mm] 0 & \vec{w}_{\vec{k}} \otimes \vec{w}^\dag_{\vec{k}} \mate.   
\ee
From the symmetric structure of the spectrum and the presence of two flat bands, we can simply infer the non-zero ones as (\change{see Fig.~$1$ in the main text})
\begin{widetext}
\be
\begin{split}
	\pm \sqrt{\vec{w}^\dag_{\vec{k}} \cdot \vec{w}_{\vec{k}}} & = \pm \sqrt{ \abs{1 + \rme{\rmi\vec{k} \cdot \vec{a}_1} }^2 + \abs{  1 + \rme{\rmi\vec{k} \cdot \vec{a}_2} }^2 + \abs{  1 + \rme{\rmi\vec{k} \cdot( \vec{a}_1 - \vec{a}_2)}}^2 }\\ 
	& = \pm \sqrt{2} \sqrt{3 + \cosa{ \vec{k} \cdot \vec{a}_1} + \cosa{ \vec{k} \cdot \vec{a}_2} + \cosa{ \vec{k} \cdot (\vec{a}_1 - \vec{a}_2)}}.
\end{split}
\ee
\end{widetext}

Choosing the reciprocal lattice vectors as 
\be
	\vec{a}_1^\ast = \frac{4\pi}{\sqrt{3}} \lt \cos \frac{\pi}{6}, -\sin \frac{\pi}{6} \rt^\intercal \mand \vec{a}_2^\ast = \frac{4\pi}{\sqrt{3}} \lt 0, 1 \rt^\intercal
\ee
the first Brillouin zone $\mal{B}$ is an hexagon in $\vec{k}$ space identified by the conditions
\be
\begin{split}
	\lt \abs{\vec{k} \cdot \vec{a}_1^\ast} \leq \ha \abs{\vec{a}_1^\ast}^2 \rt  \,\cap\, \lt \abs{\vec{k} \cdot \vec{a}_2^\ast} \leq \ha \abs{\vec{a}_2^\ast}^2  \rt \,\cap\,  \\
	\cap  \lt \abs{\vec{k} \cdot (\vec{a}_1^\ast - \vec{a}_2^\ast)} \leq \ha \abs{(\vec{a}_1^\ast - \vec{a}_2^\ast)}^2 \rt.
	\label{eq:conditions}
\end{split}
\ee
%where the $\vec{k}$s take values as shown in Eq.~\eqref{eq:wavevec}. The bands are displayed in Fig.~\ref{fig:triband}.
%\begin{figure*}[t!]
%    \centering
%    \subfloat{\includegraphics[width=0.45\textwidth]{Triband1.pdf}} 
%    \subfloat{\includegraphics[width=0.45\textwidth]{Triband2.pdf}}
%    \caption{Two views of the bands in the first Brillouin zone of the triangular lattice. The middle red one is comprised of two degenerate flat bands. The upper (yellow) and lower (blue) bands are specular and gapped.}
%\label{fig:triband}
%\end{figure*}

%%%%%%%%%%%%%%%%%%%%%%%%%%%%%%%%%%%%%%%%%%%%%%%%%%%%%
%%%%%%%%%%%%%%%%%%%%%%%%%%%%%%%%%%%%%%%%%%%%%%%%%%%%%
%%%%%%%%%%%%%%%%%%%%%%%%%%%%%%%%%%%%%%%%%%%%%%%%%%%%%
%%%%%%%%%%%%%%%%%%%%%%%%%%%%%%%%%%%%%%%%%%%%%%%%%%%%%

\subsection{Example: the honeycomb lattice}

The honeycomb lattice is a triangular Bravais lattice with primitive lattice vectors
\be
	\vec{a}_1 = a \lt 1,0   \rt^\intercal \mand \vec{a}_2 = a \lt \cos \frac{\pi}{3}, \sin \frac{\pi}{3}   \rt^\intercal ,
\ee
where the lattice spacing $a$ is $\sqrt{3}$ times the edge of the hexagons, plus a basis of two vectors
\be
	\vec{b}_1 = 0 \mand \vec{b}_2 = \frac{2\vec{a}_2 - \vec{a}_1}{3}. 
\ee
In the synthetic lattice, this gives rise to a structure with a basis of $5$ elements: $2$ one-excitation sites ($n_1 = 2$) and $3$ pair ones ($n_2 = 3$), which we choose as follows:
\begin{itemize}
	\item[$\ket{\mu_1}$]: a one-excitation site at $\vec{b} = 0$.
	\item[$\ket{\mu_2}$]: a one-excitation site at $\vec{b} = \frac{2\vec{a}_2 - \vec{a}_1}{3}$.
	\item[$\ket{\nu_1}$]: a pair site at $\vec{b} = \frac{2\vec{a}_2 - \vec{a}_1}{6}$.
	\item[$\ket{\nu_2}$]: a pair site at $\vec{b} = \frac{2\vec{a}_1 - \vec{a}_2}{6}$.
	\item[$\ket{\nu_3}$]: a pair site at $\vec{b} = -\frac{\vec{a}_1 + \vec{a}_2}{6}$.
\end{itemize}
We thus see that the $\wt{C}_{-\vec{k}}$ are $2\times 3$ matrices and that there is at least one flat zero-energy band. The matrix elements can be identified column by column as follows:
\begin{itemize}
	\item[$\ket{\nu_1}$]: From $\ket{\nu_1}$ one can jump to $\ket{\mu_1}$ or to $\ket{\mu_2}$ remaining in the same Bravais lattice site.
	\item[$\ket{\nu_2}$]: From $\ket{\nu_2}$ one can jump to $\ket{\mu_1}$ in the same site or to $\ket{\mu_2}$ changing site by $\vec{j} = \vec{a}_1 - \vec{a}_2$ ($\Rightarrow \rme{i\vec{k}\cdot (\vec{a}_1 - \vec{a}_2) }$).
	\item[$\ket{\nu_3}$]: From $\ket{\nu_3}$ one can jump to $\ket{\mu_1}$ in the same site or to $\ket{\mu_2}$ changing site by $\vec{j} = - \vec{a}_2$ ($\Rightarrow \rme{- i\vec{k} \cdot \vec{a}_2}$).
\end{itemize}
Hence,
\be
	\wt{C}_{-\vec{k}} = \matb{ccc} 1 & 1  & 1 \\ 1 & \rme{\rmi\vec{k}\cdot (\vec{a}_1 - \vec{a}_2) } & \rme{- \rmi\vec{k} \cdot \vec{a}_2}      \mate \equiv\matb{c} \vec{w}_{\vec{k},1}^\dag \\ \vec{w}_{\vec{k},2}^\dag  \mate,
\ee
with $\vec{w}_{\vec{k},1/2}$ three-dimensional vectors. The matrix $M_{\vec{k}}$ is thus
\be
	M_{\vec{k}} = \matb{cc|c} 0 & 0 & \vec{w}_{\vec{k},1}^\dag  \\[1.5mm] 0 & 0 & \vec{w}_{\vec{k},2}^\dag  \\[1.5mm] \hline \vec{w}_{\vec{k},1} & \vec{w}_{\vec{k},2} & 0    \mate . 
\ee
The kernel state is a five-dimensional vector $(0,0,\vec{v}_{\vec{k}})$ which satisfies $\vec{w}_{\vec{k},1/2}^\dag \cdot \vec{v}_k = 0$.

To identify the remaining non-zero bands, we again take the square of the total matrix $M_{\vec{k}}$:
\be
	M_{\vec{k}}^2 = \matb{cc|c} \vec{w}_{\vec{k},1}^\dag \cdot \vec{w}_{\vec{k},1}  & \vec{w}_{\vec{k},1}^\dag \cdot \vec{w}_{\vec{k},2}  & 0  \\[1mm]
	\vec{w}_{\vec{k},2}^\dag \cdot \vec{w}_{\vec{k},1} & \vec{w}_{\vec{k},2}^\dag \cdot \vec{w}_{\vec{k},2} & 0   \\[1mm] \hline 
	0 & 0 & \vec{w}_{\vec{k},1} \otimes \vec{w}_{\vec{k},1}^\dag + \vec{w}_{\vec{k},2} \otimes \vec{w}_{\vec{k},2}^\dag
\mate ,
\ee
where the first block is $2\times 2$ and the second one $3\times 3$. We can now diagonalize the first block to find (\change{see Fig.~$1$ in the main text})
\begin{widetext}
\be
	\lambda_{\vec{k},\pm} = \ha \lqq \lt \abs{\vec{w}_{\vec{k},1}}^2 + \abs{\vec{w}_{\vec{k},2}}^2  \rt \pm \sqrt{\lt    \abs{\vec{w}_{\vec{k},1}}^2  -  \abs{\vec{w}_{\vec{k},2}}^2  \rt^2  + 4\abs{\vec{w}_{\vec{k},2}^\dag \cdot \vec{w}_{\vec{k},1}}^2 }   \rqq,
\ee
\end{widetext}
with $\lambda_{\vec{k},\pm} \geq 0$. The four non-trivial bands will thus correspond to $\pm \sqrt{\lambda_{\vec{k},+}}$ and $\pm \sqrt{\lambda_{\vec{k},-}}$. Working out the scalar products
\be
	 \abs{\vec{w}_{\vec{k},1}}^2 = \abs{\vec{w}_{\vec{k},2}}^2 = 3 
\ee
and
\be
	\abs{\vec{w}_{\vec{k},2}^\dag \cdot \vec{w}_{\vec{k},1} } = \abs{1 + \rme{\rmi\vec{k}\cdot (\vec{a}_1 - \vec{a}_2) } + \rme{- \rmi\vec{k} \cdot \vec{a}_2}}
\ee
we obtain by substitution
\begin{widetext}
\be
\begin{split}
	\lambda_{\vec{k},\pm}  =  3\pm \abs{1 + \rme{\rmi\vec{k}\cdot (\vec{a}_1 - \vec{a}_2) } + \rme{- \rmi\vec{k} \cdot \vec{a}_2}}  
	 = 3 \pm \sqrt{3 + 2\cosa{\vec{k} \cdot (\vec{a}_1 - \vec{a}_2)} + 2\cosa{\vec{k} \cdot \vec{a}_2} + 2\cosa{\vec{k} \cdot \vec{a}_1}}.
\end{split}
\ee
\end{widetext}
From the first equality we see that the second addend is always $\leq 3$. It is $3$ only when $\vec{k} = 0$ (up to reciprocal lattice translations $\vec{G}$, see \eqref{eq:rec}). Hence, $\lambda_- (\vec{k} = 0) = 0$ is a minimum and $\lambda_{+} (\vec{k}=0) = 6$ is a maximum. The bands $\pm \sqrt{\lambda_{\vec{k},-}}$ touch at $\vec{k} = 0$ with linear dispersion. Second, the argument of the absolute value will vanish when 
\be
	\vec{k} \cdot (\vec{a}_1 - \vec{a}_2) = \pm \frac{2\pi}{3} + 2\pi n \,, \ -\vec{k}\cdot \vec{a}_2 = \pm \frac{4\pi}{3} + 2\pi m   
\ee
where the signs must be chosen consistently. Up to reciprocal lattice translations, one can choose
\be
	\vec{k} = \pm\frac{1}{3} \lt \vec{a}_2^\ast - \vec{a}_1^\ast   \rt,
\ee
identifying the points at the vertices of the hexagonal first Brillouin zone (one can verify this point lies at the boundary of two of the conditions in \eqref{eq:conditions}). Therefore, the two upper bands $\sqrt{\lambda_{\vec{k},+}}$ and $\sqrt{\lambda_{\vec{k},-}}$ touch at the vertices of the first Brilluoin zone with linear dispersion and similarly do the lower bands $-\sqrt{\lambda_{\vec{k},+}}$ and $-\sqrt{\lambda_{\vec{k},-}}$.

% ===================================================================================================================================================
% ===================================================================================================================================================
\section{Localization and scaling exponents}

In Table \ref{tab:scaling} we list the scaling exponents $\nu_i$ extracted from the localization lengths $\xi_i \sim s^{\nu_i}$, $i=1,2$ at a given disorder strength $s$ as described in Fig. 3 of the main text. For comparison, and further to the discussion in the main text, we list in the \change{second and third} column scaling exponents obtained with \emph{flat} disorder distribution, where the disorder energies $\delta V$ affecting the sites of the (synthetic) Lieb ladder are drawn from a uniform interval $[-W/2, W/2]$. The \change{second (third)} column corresponds to a situation, where only the sites corresponding to the pair states (all sites of the Lieb ladder) are affected. Finally, we list in the last column the values presented in \cite{Leykam2017}. It is apparent from the Table \ref{tab:scaling} that the scaling exponents at energies $\epsilon=1,1.8$ and $\sqrt{6}$ show a reasonable agreement corresponding to the ''out" and ''in" Anderson scalings 0 and 2.
\change{For $\epsilon = \sqrt{2}$, we observe for the "in" scaling a different behaviour between the cases listed in Table 1. The anomalous value of 4/3 appears only when the disorder acts on all sites. However, when the disorder affects only the pair sites  it corresponds to the Anderson value of 2 (both when the disorder is flat and drawn from the distribution (\ref{eq:distr0})).}
In contrast, we remark that for $\epsilon=\sqrt{2}$, for the flat distribution the result is independent on whether it acts on all or only on pair state sites. On the other hand, the values we obtain for the disorder distribution drawn from (\ref{eq:distr0}), i.e. $\nu(\epsilon=\sqrt{2}) \approx \{1.1, 1.1\}$ doesn't seem to be close to either the anomalous or the edge scaling exponents and, based solely on the present analysis, cannot be simply attributed to the disorder acting on only the pair state sites.

\begin{table*}[t!]
\begin{center}
\begin{tabular}{ | c || c | c | c | c |} \hline
  & \makecell{experimental disorder \\ $s \in [5 \cdot 10^{-6}, 5 \cdot 10^{-4}]$} & \makecell{flat disorder \\ on pair-state sites $(A_i,B_i,E_i)$ \\ $W \in [5 \cdot 10^{-2}, 1]$} & \makecell{flat disorder \\ on all sites $(A_i,B_i,C_i,D_i,E_i)$ \\ $W \in [1 \cdot 10^{-1}, 1]$} & \makecell{flat disorder \\ on all sites $(A_i,B_i,C_i,D_i,E_i)$ \\ values from Ref.~\cite{Leykam2017}}  \\ 
   \hline \hline
 $\epsilon = 1$ 		& $(0, 2.2)$	& $(0, 2.0)$	& $(0, 1.8)$	& $(0,2)$             \\ \hline  
 $\epsilon = \sqrt{2}$ 	& $(0.7, 2.2)$	& $(0.7, 2.0)$	& $(0.8, 1.4)$	& $(2/3, 4/3)$         \\ \hline
 $\epsilon  = 1.8$ 		& $(2.0, 1.9)$	& $(2.0, 1.8)$	& $(2.0, 2.0)$	& $(2,2)$          \\ \hline
 $\epsilon = 2$ 		& $(1.1, 1.1)$	& $(0.7, 1.3)$	& $(0.7, 1.3)$	& $(2/3, 4/3)$        \\ \hline
 $\epsilon = \sqrt{6}$ 	& $(0, 0.6)$	& $(0, 0.6)$	& $(0, 0.6)$	& $(0, 2/3)$       \\
 \hline
\end{tabular}
\end{center}
	\caption{Scaling exponents $\nu$ for different energies $\epsilon$ obtained from fitting the behaviour of the localization lengths $\xi$. $\xi_i \sim s^\nu$ for the second and $\xi_i \sim W^\nu$ for the third and fourth columns, see text for details. The range of $s$ and $W$ in the first row denote the interval of the disorder parameter over which the fit was performed. Values in the second column obtained for $\alpha=3$ and $N=10^6$. $N=10^6$ and $10^5$ has been used in the third and fourth column.}
	\label{tab:scaling}
\end{table*}

% ===================================================================================================================================================
% ===================================================================================================================================================
\section{Initial state preparation and evolution}

 In this section we consider the preparation of the state $\ket{\psi_{\rm loc}} = 1/\sqrt{4} \left( A_i + B_i - E_i - E_{i+1} \right)$ localized at rungs $i,i+1$ of the ladder. We assume that each atom in the ladder can be addressed individually with a laser pulse of Rabi frequency $\Omega_R$ and duration $\tau$ so that the atomic spin evolves according to 
\[
	U(\theta \equiv \Omega_R \tau) = {\rm e}^{-\rmi \frac{\theta}{2} \sigma_x} = \begin{pmatrix} \cos \frac{\theta}{2} & -\rmi \sin \frac{\theta}{2} \\
	-\rmi \sin \frac{\theta}{2} & \cos \frac{\theta}{2}
	\end{pmatrix}
	\label{eq:U}
\]
if the laser detuning $\Delta$ is set such that it is resonant with the transition of the addressed atom. If $\Delta \gg \Omega_{\rm R}$, instead, it acts trivially like an identity operator. Specifically, we will distinguish two special cases, namely $\Delta=0$ in addition to $\Delta=-V(R_0)$ corresponding to the blockade and facilitation condition respectively. The state $\ket{\psi_{\rm loc}}$ can be obtained by application of six pulses on initially all atoms in the spin-down state as $\ket{\psi_{\rm loc}} = \mathcal{F}_2(2 \pi) \mathcal{F}_4(2 \pi) \mathcal{F}_3(\pi) \mathcal{F}_2(\frac{\pi}{2}) \mathcal{B}_4(\pi) \mathcal{B}_1(\frac{\pi}{2}) \ket{\psi_{\downarrow..\downarrow}}$, where $\mathcal{B}_j(\theta),\mathcal{F}_j(\theta)$ stand for the laser pulse of area $\theta=\Omega_R \tau$ in the blockaded ($\mathcal{B}$) and facilitated ($\mathcal{F}$) regime applied at site $j=1,..,4$ labeling the effective plaquette formed by the four sites corresponding to the $i$-th and $(i+1)$-th rung of the ladder, see Eq. (\ref{eq:preparation}). Here, the first pulse creates an excitation at site 1, the second pulse then exploits the blockade mechanism to create a superposition of spin-up states at sites 1 and 4. Next, the pulse in the facilitated regime applied at site 2 creates a superposition of the form $-i \ket{{\color{red} \uparrow}} + \ket{\downarrow}$ if and only if a single nearest-neighbor is already excited, and so forth. We have omitted the global $-i$ factors in the second, and fourth lines of (\ref{eq:preparation}). In practice the choice of $\Omega_R$ is a trade-off between the need to keep the state-preparation time to a minimum (implying higher values of $\Omega_{\rm R}$) and the upper bounds imposed for
keeping the blockade and facilitation conditions preserved, see \cite{a:Ostmann_17} for details of these issues.

\begin{eqnarray}
\boxed{
\begin{matrix}
  \downarrow_1 & \downarrow_2 \\
  \downarrow_3 & \downarrow_4
 \end{matrix}
}
% ***************************************************
	& \xrightarrow[]{\mathcal{B}_1(\frac{\pi}{2})} &
-i \, \boxed{
\begin{matrix}
  {\color{red} \uparrow} & \downarrow \\
  \downarrow & \downarrow
 \end{matrix}
}
+
\boxed{
\begin{matrix}
  \downarrow & \downarrow \\
  \downarrow & \downarrow
 \end{matrix}
} 
\nonumber \\
%======================================================================
	& \xrightarrow[]{\mathcal{B}_4(\pi)} &
\phantom{-i \,} \boxed{
\begin{matrix}
  {\color{red} \uparrow} & \downarrow \\
  \downarrow & \downarrow
 \end{matrix}
}
+
\boxed{
\begin{matrix}
  \downarrow & \downarrow \\
  \downarrow & {\color{red} \uparrow}
 \end{matrix}
} 
\nonumber \\
%======================================================================
	& \xrightarrow[]{\mathcal{F}_2(\frac{\pi}{2})} &
-i\,\boxed{
\begin{matrix}
  {\color{red} \uparrow} & {\color{red} \uparrow} \\
  \downarrow & \downarrow
 \end{matrix}
}
+
\boxed{
\begin{matrix}
  {\color{red} \uparrow} & \downarrow \\
  \downarrow & \downarrow
 \end{matrix}
}
-i\,\boxed{
\begin{matrix}
  \downarrow & {\color{red} \uparrow} \\
  \downarrow & {\color{red} \uparrow}
 \end{matrix}
}
+
\boxed{
\begin{matrix}
  \downarrow & \downarrow \\
  \downarrow & {\color{red} \uparrow}
 \end{matrix}
} 
\nonumber \\
%======================================================================
	& \xrightarrow[]{\mathcal{F}_3(\pi)} &
\phantom{-i \,}\boxed{
\begin{matrix}
  {\color{red} \uparrow} & {\color{red} \uparrow} \\
  \downarrow & \downarrow
 \end{matrix}
}
+
\boxed{
\begin{matrix}
  {\color{red} \uparrow} & \downarrow \\
  {\color{red} \uparrow} & \downarrow
 \end{matrix}
}
+
\boxed{
\begin{matrix}
  \downarrow & {\color{red} \uparrow} \\
  \downarrow & {\color{red} \uparrow}
 \end{matrix}
}
+
\boxed{
\begin{matrix}
  \downarrow & \downarrow \\
  {\color{red} \uparrow} & {\color{red} \uparrow}
 \end{matrix}
} 
\nonumber \\
%======================================================================
	& \xrightarrow[]{\mathcal{F}_4(2\pi)} &
\phantom{-i \,}\boxed{
\begin{matrix}
  {\color{red} \uparrow} & {\color{red} \uparrow} \\
  \downarrow & \downarrow
 \end{matrix}
}
+
\boxed{
\begin{matrix}
  {\color{red} \uparrow} & \downarrow \\
  {\color{red} \uparrow} & \downarrow
 \end{matrix}
}
-
\boxed{
\begin{matrix}
  \downarrow & {\color{red} \uparrow} \\
  \downarrow & {\color{red} \uparrow}
 \end{matrix}
}
-
\boxed{
\begin{matrix}
  \downarrow & \downarrow \\
  {\color{red} \uparrow} & {\color{red} \uparrow}
 \end{matrix}
} 
\nonumber \\
%======================================================================
	& \xrightarrow[]{\mathcal{F}_2(2\pi)} &
\phantom{-i \,}\boxed{
\begin{matrix}
  {\color{red} \uparrow} & {\color{red} \uparrow} \\
  \downarrow & \downarrow
 \end{matrix}
}
-
\boxed{
\begin{matrix}
  {\color{red} \uparrow} & \downarrow \\
  {\color{red} \uparrow} & \downarrow
 \end{matrix}
}
-
\boxed{
\begin{matrix}
  \downarrow & {\color{red} \uparrow} \\
  \downarrow & {\color{red} \uparrow}
 \end{matrix}
}
+
\boxed{
\begin{matrix}
  \downarrow & \downarrow \\
  {\color{red} \uparrow} & {\color{red} \uparrow}
 \end{matrix}
} \nonumber \\
\label{eq:preparation}
\end{eqnarray}

Once the state $\ket{\psi_{\rm loc}}$ has been prepared, it evolves according to $H_{\rm eff}$, Eq. (\ref{eq:H eff}). We define the probability of excitation at rung $i$ is obtained as $p^\alpha_i = n^\alpha_i/\sum_{i=1}^{L} n^\alpha_i$. Here $n^\alpha_i = \bra{\psi(t)} \hat{n}^\alpha_i \ket{\psi(t)}$, $\alpha=u,l$ for the upper and lower leg of the ladder respectively.
%\begin{equation}
%	\hat{n}^\alpha_i =
%		\ket{S^\alpha_{i-1}}\bra{S^\alpha_{i-1}} + \ket{C_{i}}\bra{C_{i}} + \ket{S^\alpha_{i}}\bra{S^\alpha_{i}} + \ket{E_{i}}\bra{E_{i}}
%\end{equation}
%\begin{equation}
%	\hat{n}_i = 
%	\begin{cases}
%		\ket{A_{i-1}}\bra{A_{i-1}} + \ket{C_{i}}\bra{C_{i}} + \ket{A_{i}}\bra{A_{i}} + \ket{E_{i}}\bra{E_{i}} \\
%		\text{ upper leg} \\
%		\ket{B_{i-1}}\bra{B_{i-1}} + \ket{D_{i}}\bra{D_{i}} + \ket{B_{i}}\bra{B_{i}} + \ket{E_{i}}\bra{E_{i}} \\
%		\text{ lower leg}
%	\end{cases}
%\end{equation}
%is the projector operator on all spin configurations which contain an excitation at site $i$ and $S^u=A, S^l=B$. 
We then define the average position $\bar{x}$ and the standard deviation $\Delta x$ of the excitations as
\begin{eqnarray}
	\bar{x}^\alpha &=& \sum p^\alpha_i i \\
	\left( \Delta x^\alpha \right)^2 &=& \sum_i p^\alpha_i i^2 - \left(\bar{x}^\alpha \right)^2 = \sum_i p^\alpha_i (i-\bar{x}^\alpha)^2.
	\label{eq:Delta x}
\end{eqnarray}

% ===================================================================================================================================================
% ===================================================================================================================================================
\section{Numerical simulation of the spin dynamics}

We consider the full Hamiltonian Eq. (\ref{Eq:Hamil_full}) which we express in units of the Rabi frequency as
\begin{equation}
	\Omega^{-1} \op H = \sum_k \op \sigma_x^{(j)} + (-\tilde{V}(R_0)) n_k + \tilde{V}(R_0) \sum_{k>j} \frac{\op n_k \op n_j}{|k-j|^\alpha},
\end{equation}
where $\tilde{V}(R_0) = V(R_0)/\Omega$ (in what follows we label all dimensionless quantities by tilde). This leads to the following effective Hamiltonian on the Lieb lattice of length $L$ 
\begin{equation}
	\Omega^{-1} \op H_{\rm eff} = \op{\tilde{H}}_0 \otimes \mathds{1}_L + \op{\tilde{H}}_0^{\rm dis} + \left[ \op{\tilde{H}}_1 \otimes \op G_L + {\rm H.c.} \right],
	\label{eq:H eff}
\end{equation}
expressed in the basis
\begin{equation}
	\{A_1,..,A_L,B_1,..,E_L\},
	\label{eq:basis}
\end{equation}
where
\begin{equation}
	\op{\tilde{H}}_0 =
 \begin{pmatrix}
  	 0 & 0 & 1 & 0 & 0 \\
     0 & 0 & 0 & 1 & 0 \\
     1 & 0 & 0 & 0 & 1 \\
     0 & 1 & 0 & 0 & 1 \\
     0 & 0 & 1 & 1 & 0
 \end{pmatrix},
\end{equation}
$\op H_1$ is a $5 \times 5$ matrix with only non-zero entries $(\op H_1)_{1,3}=(\op H_1)_{2,4}=1$, $(G_L)_{ij} = \delta_{i, j-1}$ is a $L \times L$ matrix with ones on the first upper diagonal and 
\begin{eqnarray}
	\op{\tilde{H}}_0^{\rm dis} = {\rm diag} & & \left( \tilde{\delta}_{A_1},...,\tilde{\delta}_{A_{L-1}},\tilde{\delta}_{A_{L}}=0, \right. \nonumber \\
									  & & \tilde{\delta}_{B_1},...,\tilde{\delta}_{B_{L-1}},\tilde{\delta}_{B_{L}}=0, \nonumber \\
									  & & \tilde{\delta}_{C_1}=0,...,\tilde{\delta}_{C_{L}}=0, \nonumber \\
									  & & \tilde{\delta}_{D_1}=0,...,\tilde{\delta}_{D_{L}}=0, \nonumber \\
									  & & \left. \tilde{\delta}_{E_1},...,\tilde{\delta}_{E_{L}} \right)
	\label{eq:disorder}
\end{eqnarray}
is a $5L \times 5L$ diagonal disorder matrix, where we impose open boundary conditions by requiring that $\tilde{\delta}_{A_L} = \tilde{\delta}_{B_L}=0$, since spin configurations corresponding to the $A_L,B_L$ basis elements
are missing; in fact, these would be pair states occupying a site on the $L$-th rung and one on the (non-existent) ($L+1$)-th. Analogously we enforce the OBC in the coupling matrix by setting all Hamiltonian elements corresponding to the $L$-th and $2L$-th rows and columns to 0. Here, $\tilde{\delta}_{\Xi_j} = \Omega^{-1} \left(V(d_{\Xi_j})-V(R_0) \right)$, where $\Xi_j=A_j,..,E_j$ and $d_{\Xi_j}$ is a shorthand for the spin separation in the given configuration $\Xi_j$. We note that since configurations $C,D$ correspond to single spin excitation, the associated disorder is vanishing by definition, $\tilde{\delta}_{C_j} = \tilde{\delta}_{D_j}=0, \; \forall j$. The disorder energies $\tilde{\delta}_{\Xi_j}$ are generated from first drawing a specific realization of atomic positions at each site of the lattice in all three spatial directions with isotropic Gaussian distribution of width $s$.

We then exactly evolve an initial state
\begin{equation}
	\ket{\psi_0} = \sum_{j=1}^{5L} c_j \ket{b_j},
\end{equation}
as $\ket{\psi(t)} = {\rm exp}\left[ -\rmi t \op H_{\rm eff}\right] \ket{\psi_0}$, where $b_j$ are the elements of the basis (\ref{eq:basis}) [strictly speaking there are only $5L-2$ non-trivial elements due to the OBC].

We note that the result of the evolution depends on two independent parameters, $s$ and the ratio $V(R_0)/\Omega$, where the Rabi frequency should further satisfy $\Omega \ll V(2 R_0)$ for the effective Hamiltonian (\ref{eq:H eff}) to be valid. In Fig. 5a we present the results of the simulation analogous to that performed in Fig. 4, showing $\Delta x$ in the $s - \Omega/V(R_0)$ plane. We observe that the maximum of $\Delta x$ as a function of the disorder gets shifted towards higher disorder strength as $\Omega$ is increased. 

%First we note that in the dimensionless formulation, the off-diagonal hopping elements in $H_{\rm eff}$ are of a fixed amplitude 1. 
The dependence of $\Delta x$ in Fig. 5a can be intuitively understood as follows. Smaller values of $\Omega/V(R_0)$ correspond to larger diagonal disorder elements $\tilde{\delta}$. Since it is the disorder which couples the flat and dispersive bands, the smaller the $s$, the smaller the $\Omega$ that is sufficient to cause the excitation hopping and thus the increase in $\Delta x$.
As $s$ is increased, Anderson localization becomes more and more relevant and, correspondingly, the localization length at $\epsilon = 0$ shrinks. Eventually, the state becomes capable of propagating over distances comparable to the localization length. Further increasing $s$ then reduces this scale, corresponding to the decrease in $\Delta x$.
Clearly, by increasing $\Omega/V(R_0)$ the hopping amplitude becomes more relevant with respect to the typical energy shifts and the localization length is thus increased. Higher values of $s$ are then required to localize the state again.
In Fig. 5b,c we show a comparison between the exact evolution according to the full Hamiltonian (\ref{Eq:Hamil_full}), dashed line, and $H_{\rm eff}$, solid line. As expected, the predictions of the two models show an agreement in the regime where $\Omega \ll V(2 R_0)$, Fig.5c ($V(R_0)/\Omega=200$). On the other hand for larger $\Omega$, the two models start to differ as shown in Fig. 5b ($V(R_0)/\Omega=20$). 

\begin{figure}

	      \includegraphics[width=0.5\textwidth]{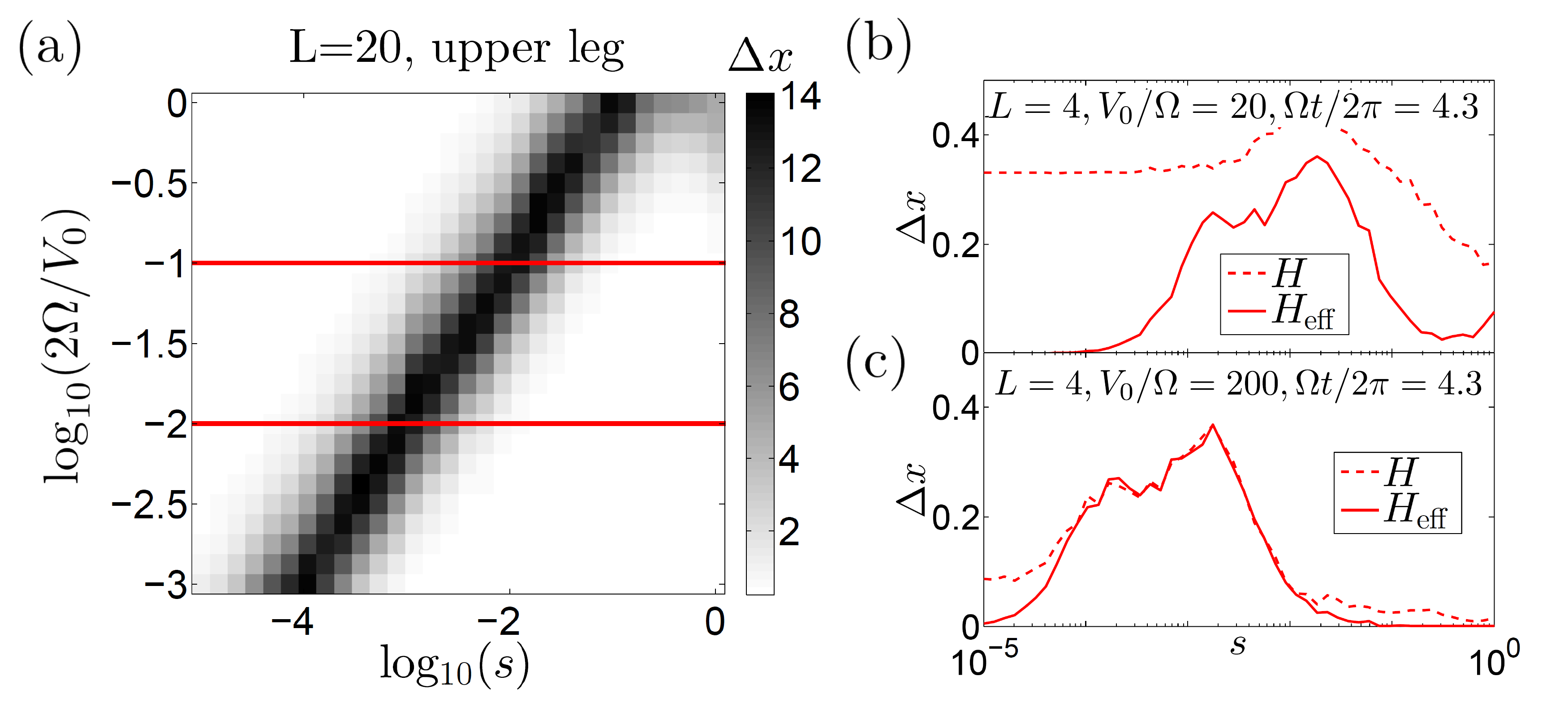}
		\caption{(a) \change{Width} $\Delta x$, Eq. (\ref{eq:Delta x}), of the excitation positions in the $s-\Omega/V(R_0)$ plane. Here, $\Delta x$ was obtained by evolving the initial state $\ket{\psi_{\rm loc}}$ located at rungs 10 and 11 in the middle of the ladder of length $L=20$ by the effective Hamiltonian $H_{\rm eff}$. The two red solid lines correspond to a cut for fixed values of $\Omega/V(R_0)$, $\Omega/V(R_0)=1/20$ (upper line) and $\Omega/V(R_0)=1/200$ (lower line). (b) Comparison between the evolution of $\ket{\psi_{\rm loc}}$ generated by $H$, Eq. (\ref{Eq:Hamil_full}), dashed line and $H_{\rm eff}$, solid line in a ladder of $L=4$ and for $\Omega/V(R_0)=1/20$. (c) Same as (b) with $\Omega/V(R_0)=1/200$. Here we have used $\Omega t/2 \pi = 4.3$ for each respective $\Omega$ and averaged over 100 disorder realizations. 
		}
 \label{Fig:Dx_scan}
   
\end{figure}


% ===================================================================================================================================================
% ===================================================================================================================================================

\end{appendix}

\end{document}